\documentclass[aps,prb,preprint,groupedaddress,showpacs,amsmath,amssymb]{revtex4}

\usepackage{graphicx}
\usepackage{dcolumn}
\usepackage{bm}
\usepackage{hyperref}

\begin{document}

\title{Highly responsive ground state of  PbTaSe$_2$: structural phase transition and evolution of superconductivity under pressure}

\author{Udhara Kaluarachchi$^{1,2}$, Yuhang Deng$^3$, Matthew F. Besser$^1$, Kewei Sun$^1$,  Lin Zhou$^1$, Manh Cuong Nguyen$^{1}$,  Zhujun Yuan$^4$, Chenglong Zhang$^4$, James S. Schilling$^3$, Matthew J. Kramer$^1$,  Shuang Jia$^{4}$, Cai-Zhuang Wang$^{1}$, Kai-Ming Ho$^{1,2}$,  Paul C. Canfield$^{1,2}$, and Sergey L. Bud'ko$^{1,2}$ }
\affiliation{$^{1}$Ames Laboratory, US DOE,  Iowa State University, Ames, Iowa 50011, USA}
\affiliation{$^{2}$Department of Physics and Astronomy, Iowa State University, Ames, Iowa 50011, USA}
\affiliation{$^{3}$Department of  Physics, Washington University, St. Louis, Missouri 63130, USA}
\affiliation{$^{4}$International Center for Quantum Materials, School of Physics, Peking University, Beijing 100871, China}

\date{\today}

\begin{abstract}

 Transport and magnetic studies of PbTaSe$_2$ under pressure suggest existence of two superconducting phases with the low temperature phase boundary at $\sim 0.25$ GPa  that is defined by a very sharp, first order, phase transition. The first order phase transition line can be followed via pressure dependent resistivity measurements, and is found to be near 0.12 GPa near room temperature. Transmission electron microscopy and x-ray diffraction at elevated temperatures confirm that this first order phase transition is structural and occurs at ambient pressure near $\sim 425$ K. The new, high temperature / high pressure phase has a  similar crystal structure and slightly lower unit cell volume relative to the ambient pressure, room temperature structure.  Based on first-principles calculations this structure is suggested to be obtained by shifting the Pb atoms from the $1a$  to $1e$ Wyckoff position without changing the positions of Ta and Se atoms. PbTaSe$_2$ has an exceptionally pressure sensitive, structural phase transition with $\Delta T_s/\Delta P \approx  - 1700$ K/GPa near 4 K, this first order transition causes an $\sim 1$ K ($\sim 25 \%$) step - like decrease in $T_c$ as pressure is increased through 0.25 GPa.

\end{abstract}


\maketitle

\section{Introduction}

Although PbTaSe$_2$ was discovered several decades ago, \cite{epp80a,epp81a} its electronic structure and physical properties have only been studied in  detail over the past few years. \cite{bia16a,ali14a,wan15a,wan16a,pan16a,zha16a} Structurally, PbTaSe$_2$ can be thought of  as alternating stacking of hexagonal TaSe$_2$ and Pb layers with the $P\bar{6}m2$ space group. The crystal structure of PbTaSe$_2$ is non-centrosymmetric. \cite{epp80a} Initially, only the values of resistivity and Hall coefficient at 300 K of a pressed powder pellet of PbTaSe$_2$ and resistive onset of superconductivity at $T_c = 6.5$ K were reported. \cite{epp81a}

Recently PbTaSe$_2$ was identified as a topological, nodal semimetal with strong spin-orbit coupling. \cite{bia16a} Its superconducting transition temperature in polycrystalline samples \cite{ali14a,wan15a} and in single crystals \cite{wan16a,pan16a,zha16a} based on thermodynamic and transport measurements was established to be $\sim 3.8$ K. Thermal conductivity and  London penetration depth measurements \cite{wan16a,pan16a} suggested a nodeless superconducting gap structure for this material. Low temperature magnetoresistance was found
 to be relatively high, anisotropic, and sublinear in magnetic field; the anisotropy of the upper critical field, $H_{c2}$, from the resistivity measurements was reported to be temperature dependent, with the values of $\gamma = H_{c2}^{ab}/H_{c2}^c = 2 - 4$. \cite{zha16a}

The complex, non-trivial band structure of  PbTaSe$_2$ \cite{bia16a,ali14a,zha16a} suggests possible sensitivity of its physical properties to applied pressure. Indeed, a non-monotonic, V-shaped, pressure dependence of $T_c$ was observed in a polycrystalline sample of PbTaSe$_2$ that had a relatively broad resistive superconducting transition. \cite{wan15a} This behavior of $T_c(P)$ was suggested to result from a Lifshitz transition under pressure.

Non-monotonic behavior of $T_c$ as a function of pressure has been observed in a number of materials, including elements. \cite{hat56a,jen58a,chu68a} A Lifshitz transition (a change of the Fermi surface topology) \cite{lif60a} has been invoked to explain such evolution of $T_c$ with pressure. \cite{mak65a} Recently, for complex superconductors, other possible causes for non-monotonic behavior of $T_c$ under pressure were discussed and studied, e.g. crossing long range magnetic order phase lines in $T - P$ phase diagrams \cite{has12a,bud14a,bud15a,ben10a,ben12a,kal16a} or pressure-induced changes in superconducting pairing symmetry or gap structure. \cite{taf13a,ter14a,tau14a,taf15a} 

To clarify the intrinsic $T_c(P)$ behavior of PbTaSe$_2$ under pressure and to better address the physics associated with this behavior, in this work we perform measurements of in-plane resistivity of high quality PbTaSe$_2$ single crystals under pressures up to $\sim 1.5$ GPa in zero and applied magnetic field. In addition to $T_c(P)$, and pressure dependence of the normal state resistivity, these measurements allow us to follow the evolution of the upper critical field, $H_{c2}^c(T)$ near the $T_c(H = 0)$ (that was instrumental in e.g. studies of KFe$_2$As$_2$ and FeSe \cite{tau14a,taf15a,kal16a}) and of the low temperature, normal state magnetoresistance.

Additionally, motivated by the low temperature results discussed below, ambient pressure x-ray diffraction and transmission electron microscopy  measurements at elevated temperatures were performed. Experimental studies were complimented by the first-principles calculations of the stability of different crystallographic phases under pressure.

\section{Experimental and computational methods}

\subsection{Experimental details}

PbTaSe$_2$ single crystals were grown by chemical vapor transport,  using PbCl$_2$ as a transport agent. More details about the synthesis are presented in Ref. \onlinecite{zha16a}. The quality of the samples was attested by rather high residual resistivity ratios, $RRR = \rho_{300K}/\rho_{4K} \sim 115-120$ and sharp superconducting transitions in zero field.  The electrical contacts for standard 4-probe ac resistivity measurements were made using Pt wires and combination of Du Pont 4929N silver paste and Epo-Tek H20E silver epoxy. The current was flowing in the $ab$ plane and magnetic field was applied along the $c$-axis. 

The resistivity measurements at ambient and high  pressure were performed in a Quantum Design Physical Property Measurement System (PPMS-9). For resistivity measurements under pressure a Be-Cu/Ni-Cr-Al hybrid piston cylinder pressure cell similar to that used in Ref. \onlinecite{bud84a} was used. A 40 : 60 mixture of light mineral oil and n-pentane was used as a pressure medium. This medium solidifies at room temperature at $P \sim 3.5$ GPa \cite{bud84a,tor15a}, well above the pressure range used in this work. The pressure at room temperature was evaluated using manganin resistive gauge, whereas at low temperatures the superconducting transition of pure Pb \cite{eil81a} was used to determine pressure. Additionally, low-field dc magnetization under pressure down to 1.8 K was measured in a Quantum Design Magnetic Property Measurement System (MPMS-5) SQUID magnetometer using a commercial, HMD, Be-Cu piston-cylinder pressure cell \cite{hmd}. In these measurements Daphne oil 7373 was used as a pressure medium and superconducting Pb as a low-temperature pressure gauge \cite{eil81a}. For magnetization measurements a stack of single crystals was oriented with $H \| c$.

An additional set of resistivity measurements was performed under He - gas pressure. Four-point ac electrical resistivity measurements were carried out simultaneously on two PbTaSe$_2$ crystals with approximate dimensions 0.5$\times$0.1$\times$0.05 mm$^3$ to hydrostatic pressures as high as 0.37 GPa in a He-gas high-pressure system. An excitation current of 1 mA (rms) at 17 Hz was applied using a Keithley 6221 constant ac/dc current source and the small voltage detected by a Stanford Research SR830 digital lock-in amplifier. Experiments were carried out both at constant temperature with varying pressures or at constant pressure with varying temperatures. A Janis SuperVariTemp cryostat was used to reach temperatures below the ambient.

To generate hydrostatic pressure the samples were placed in the 7 mm diameter bore of a Be-Cu  high-pressure cell (Unipress, Warsaw) connected to a three-stage Harwood compressor system. He gas from the compressor is fed into the pressure cell via a 3 mm OD / 0.3 mm ID Be-Cu capillary tube. A calibrated digital manganin gauge at ambient temperature (Harwood model DJ-320/42) accurately determines the pressure. A sizable ``dead volume" at ambient temperature reduces the decrease of He-gas pressure on cooling from ambient to low temperatures. 

High temperature x-ray diffraction (XRD) was performed using a Panalytical X'Pert Pro XRD system with an Anton Paar HTK-1200N furnace under flowing  helium after evacuating and backfilling the system with He. Larger single crystals were ground to a few tens of $\mu$m but retained their micaceous morphology hence the XRD only exhibited (00l) reflections. Copper and cobalt radiation was used. Heating rates were either 3 K/min or 5 K/min.

Since the XRD was not able to resolve changes of the {\it in-plane} lattice with heating, additional studies using transmission electron microscopy (TEM) at different temperatures  using a FEI Tecnai G2 F200 instrument operating at 200 kV were performed. For these studies a single crystal was thinned via Ar ion milling. The thin region was co-planar with the basal planes, providing an orthogonal view of the lattice expansion on heating compared to the XRD results, i.e., the (hk0) reflections. The in situ heating/cooling was performed on a Gatan heating stage up to $\sim 500$ K. Continuous recording of the selected area diffraction was obtained on cooling. 

\subsection{First-principles calculations method}

The first-principles density functional theory (DFT) \cite{koh65a} calculations were performed using the Vienna Ab-Initio Simulation Package (VASP) \cite{kre96a} with projector-augmented wave (PAW) pseudopotential method  \cite{blo94a,kre99a} and plane wave basis. The generalized-gradient approximation parameterized for solids (PBEsol) \cite{per08a} was used for the exchange-correlation energy functional. Spin-orbit coupling was included in the calculations. The energy cutoff is 320 eV and the Monkhorst-Pack's scheme \cite{mon76a} was  used for Brillouin zone sampling with a $k$-point mesh of $12 \times 12 \times 4$ for the ground state $P\bar{6}m2$ structure and equivalent $k$-point meshes for other structures. All crystal structures were fully relaxed until the forces on each atom were smaller than 0.01 eV/\AA~and external pressure was smaller than 0.1 GPa.

\section{Results}
\begin{figure}
\begin{center}
\includegraphics[angle=0,width=120mm]{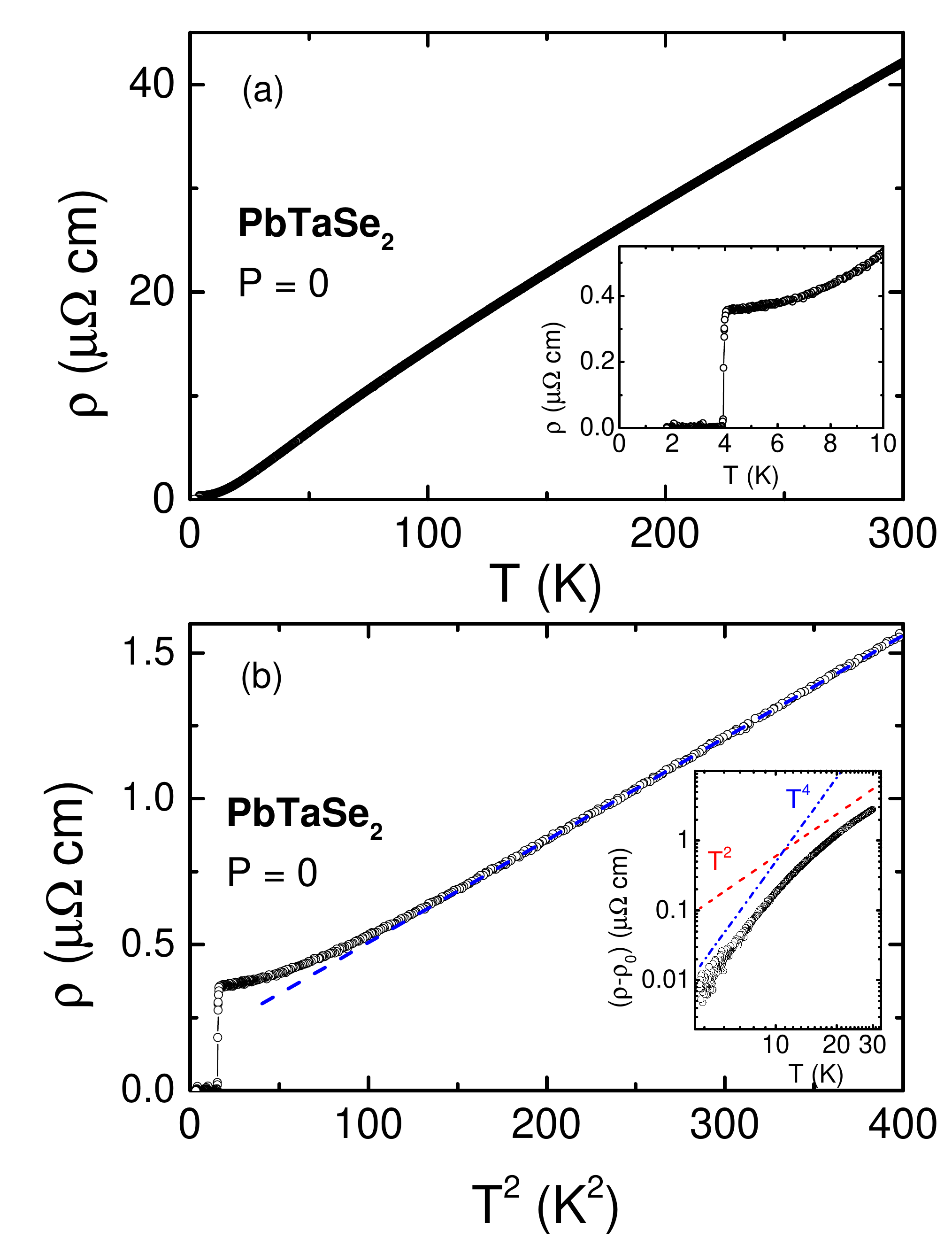}
\end{center}
\caption{(Color online) (a) Zero field, ambient pressure, in-plane  resistivity of PbTaSe$_2$. Inset: low temperature part of the data showing the superconducting transition.  (b) Low temperature resistivity of PbTaSe$_2$ at ambient pressure plotted as $\rho$ vs $T^2$. The dashed line is a guide for the eye. Inset: {\it log - log} plot of the low temperature resistivity after subtraction of the residual resistivity $\rho_0$. Lines show $T^2$ and $T^4$ behavior. } \label{F1}
\end{figure}

Ambient pressure resistivity data  (Fig. \ref{F1}a) are grossly consistent with those reported in Refs. \onlinecite{wan16a,pan16a,zha16a}. Although there is a region of $\rho = \rho_0 +AT^2$ behavior in resistivity, it does not persist to the temperatures close to the superconducting transition, where the behavior with temperature to higher power  (close to 4) was observed (Fig. \ref{F1}b). The upper critical field, $H_{c2}(T)$, was determined from electrical transport and magnetization measurements (see Appendix A for details). Here, as well as in the literature,  \cite{zha16a} there is an apparent discrepancy between the results of thermodynamic and transport measurements. 

\begin{figure}
\begin{center}
\includegraphics[angle=0,width=100mm]{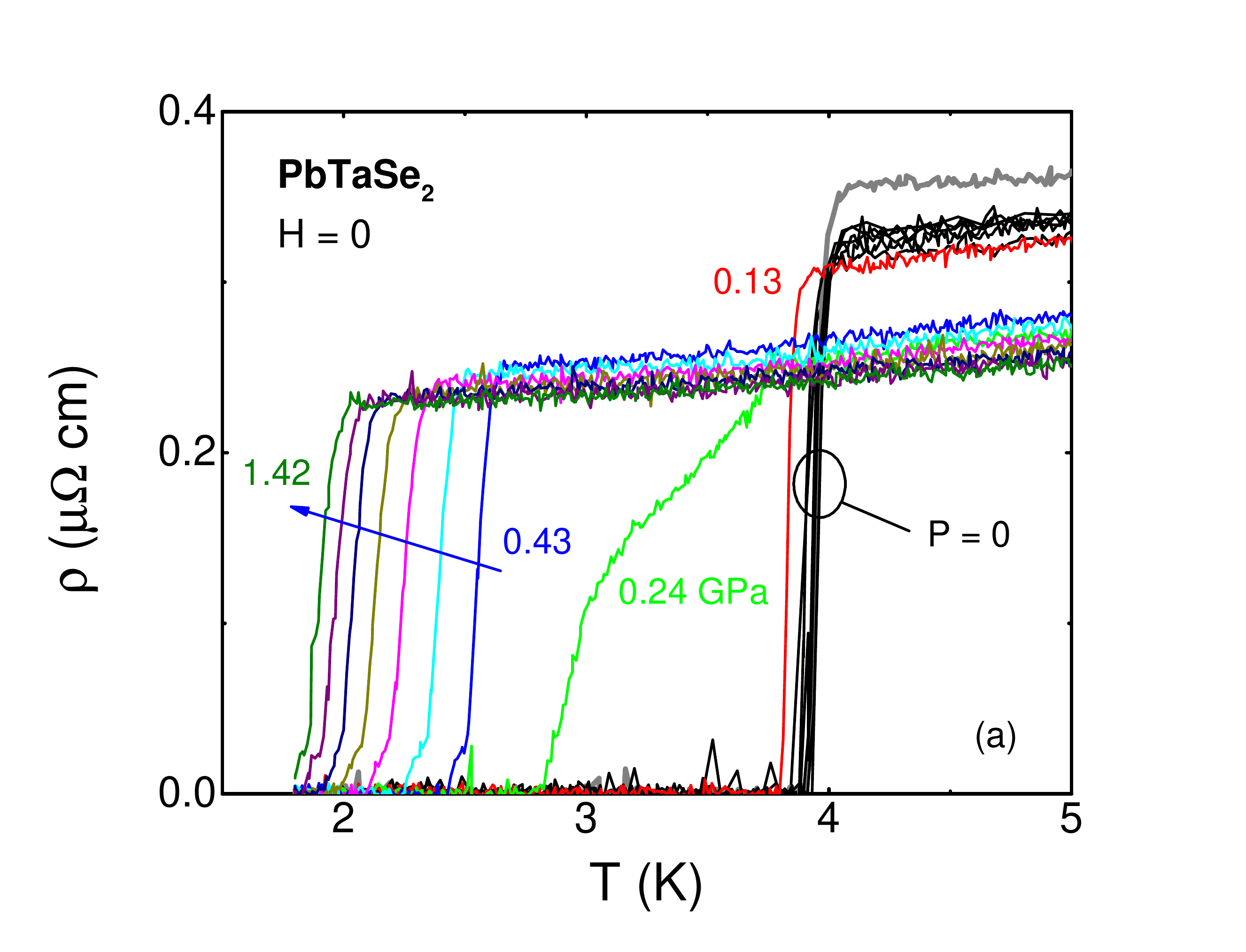}
\includegraphics[angle=0,width=100mm]{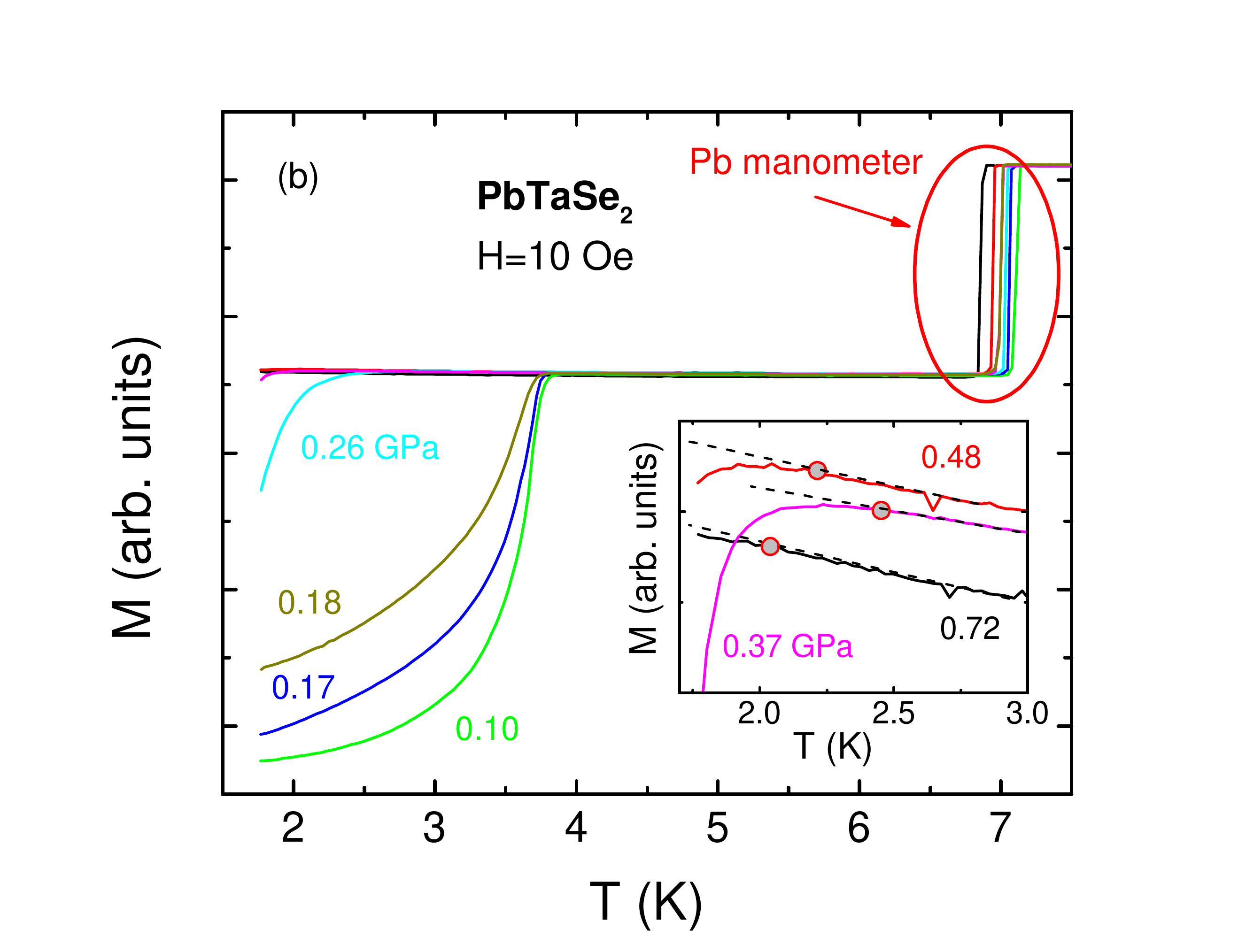}
\end{center}
\caption{(Color online) Low temperature resistivity (a) and magnetization (b) as a function of pressure. Numbers near the data are low temperature value of pressure in GPa. Several resistivity curves at $P = 0$ correspond to the pressure runs that have slightly different room temperature pressure values but result in the same $P = 0$ low temperature value as measured by Pb manometer. For magnetization measurements a deviation from linear, normal state magnetization (indicated by circle) was taken as a $T_c$ criterion, see inset to the panel (b). } \label{F3}
\end{figure}

\begin{figure}
\begin{center}
\includegraphics[angle=0,width=120mm]{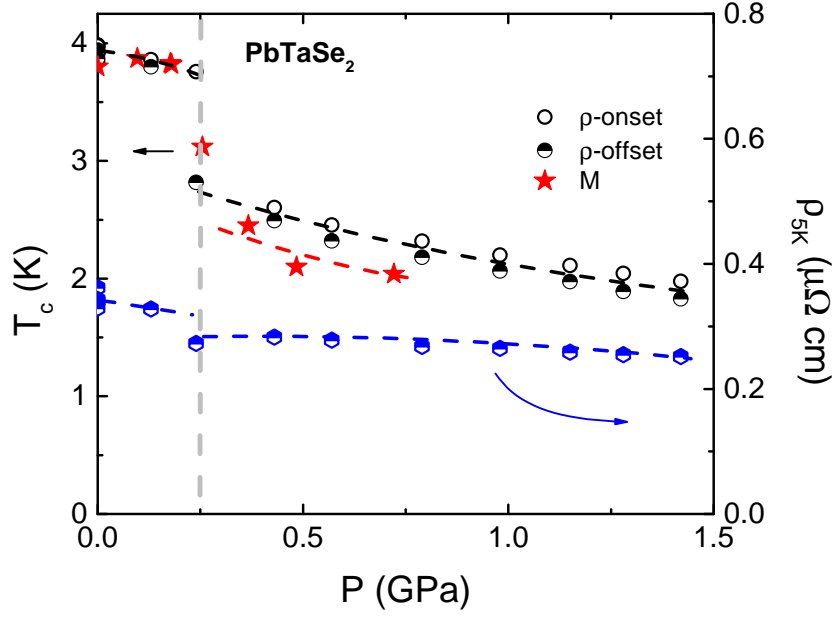}
\end{center}
\caption{(Color online) $T_c$  as determined from resistivity and magnetization measurements (left Y-axis), and normal state resistivity at 5 K (right Y-axis) as a function of pressure. Dashed vertical line shows $P = 0.25$ GPa location. Other dashed lines are guides for the eye.} \label{F4}
\end{figure}

Turning to superconducting properties under pressure, the evolution of the superconducting transition under pressure, as measured by resistivity and low field magnetization, is presented in Fig. \ref{F3}. In the resistivity measurements the transitions at all pressures, other than 0.24 GPa, are sharp. From both measurements it appears that the $T_c(P)$  has a step-like behavior. Indeed, as it is seen in Fig. \ref{F4}, both measurements result in consistent data, and a clear, sharp step in $T_c(P)$ is observed at about 0.25 GPa.  The broad, two-step-like resistive transition at 0.24 GPa corresponds to this apparent phase boundary. Normal state resistivity at 5 K, just like $T_c$, has a step-like behavior (Fig. \ref{F4}). The initial slope of transport $H_{c2}(T)$ (see Appendix B) has a step-like change at $P \approx 0.25$ GPa as well.

\begin{figure}
\begin{center}
\includegraphics[angle=0,width=120mm]{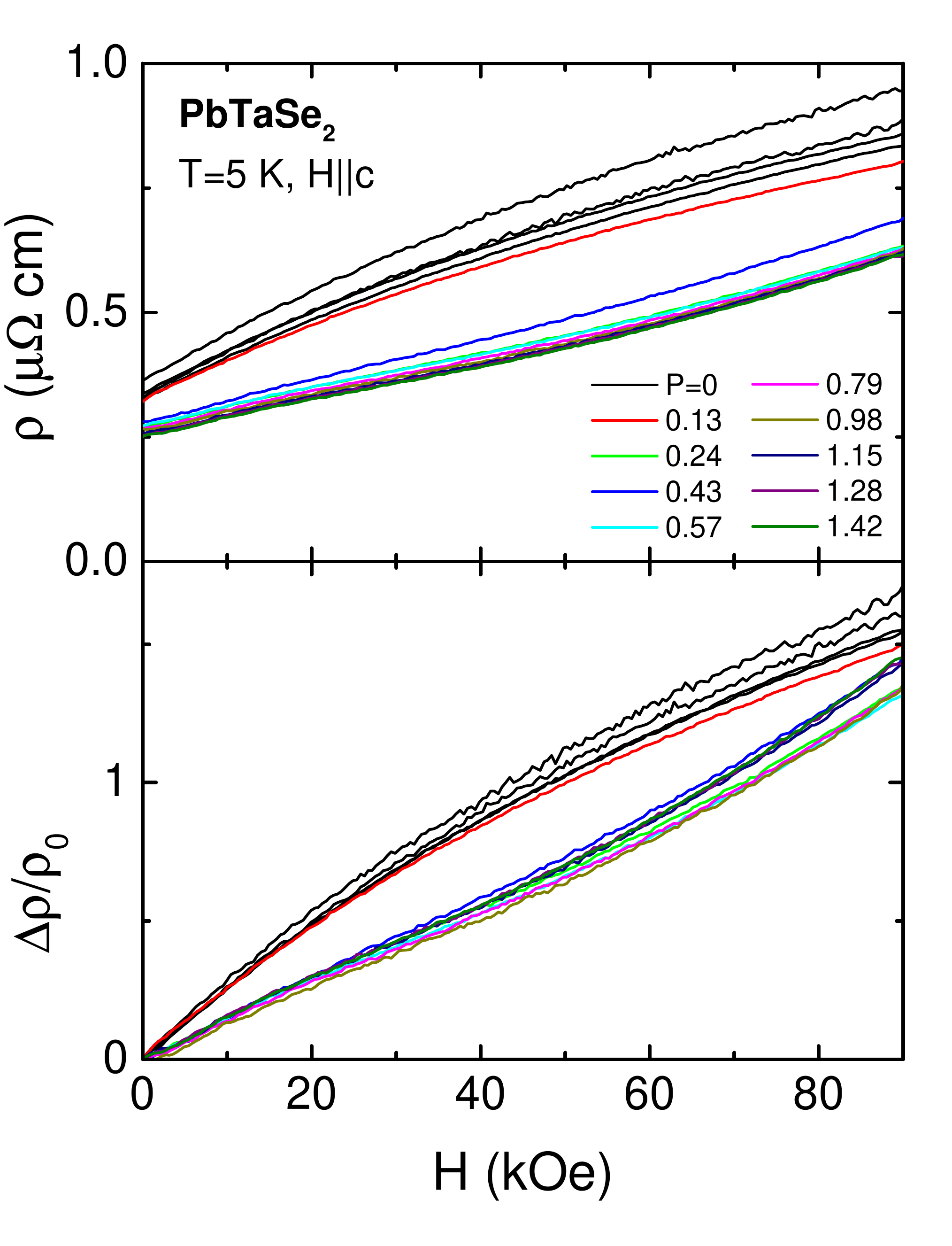}
\end{center}
\caption{(Color online) Low temperature, $T = 5$ K, field dependent magnetoresistance at different pressures (listed in the upper panel in units of GPa), plotted as  $\rho(H)$ and $\Delta \rho/\rho_0(H)$. Note: for both panels the four P = 0 data sets, as well as the $P = 0.13$ GPa data set form an upper manifold, and the $P \geq 0.24$ GPa data sets form a lower manifold.} \label{F7}
\end{figure}

Furthermore, the normal state properties show discontinuities at $P \approx 0.20-0.25$ GPa. The field dependent magnetoresistance data (Fig. \ref{F7}) clearly fall on two different manifolds, for $P\leq 0.13$ GPa and $P \geq 0.24$ GPa, that have different functional dependencies of $\Delta \rho/\rho_0$. 

\begin{figure}
\begin{center}
\includegraphics[angle=0,width=120mm]{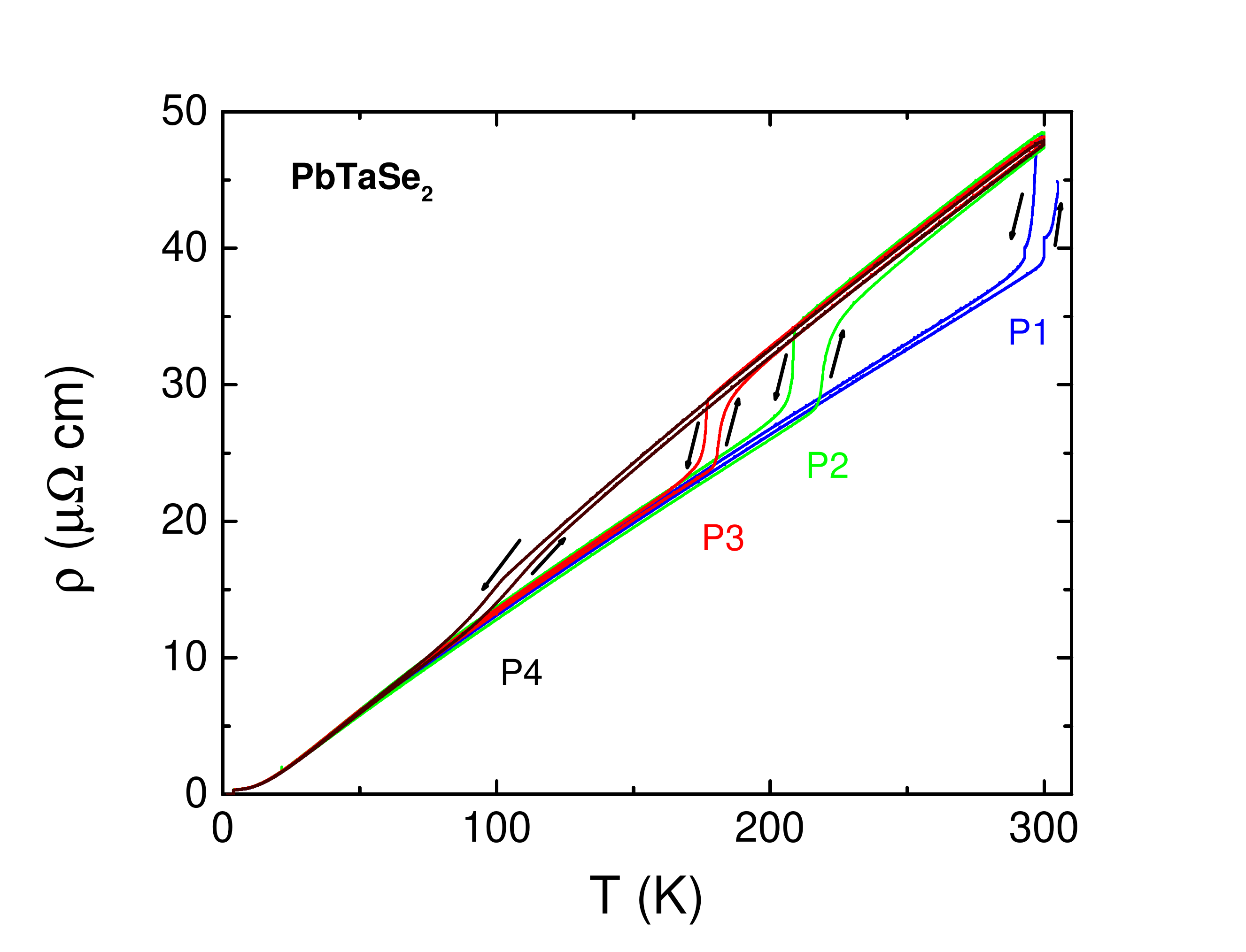}
\end{center}
\caption{(Color online) Temperature-dependent resistivity measured on warming and on cooling at four different, small pressures in the piston-cylinder cell, 0.01 GPa $<  P_1 < P_2 < P_3 < P_4 <$ 0.25 GPa (see text for more details).} \label{F9}
\end{figure}

A careful look at zero field, temperature-dependent resistivity data taken in the piston-cylinder cell at small, from 0.05 to 0.39 GPa at room temperature, pressures (Fig. \ref{F9}) reveals a clear, hysteretic in temperature, sharp transition. Although qualitatively these data point to an additional phase line, probably of the structural transition, that has a very steep pressure dependence, a quantitative analysis of these data is hindered by an experimental issue associated with a use of piston-cylinder pressure cells over an extended temperature range. Due to differential thermal contraction of the materials of the cell and the medium, the pressure inside the cell decreases on cooling \cite{tho84a}. This pressure drop depends on multiple factors, including the cell materials and design, medium and the pressure range. At low pressures, even when using manganin as a room temperature pressure gauge and Pb as a low temperature pressure gauge, the evaluation of pressure at intermediate temperatures has substantial error bars. 

\begin{figure}
\begin{center}
\includegraphics[angle=0,width=100mm]{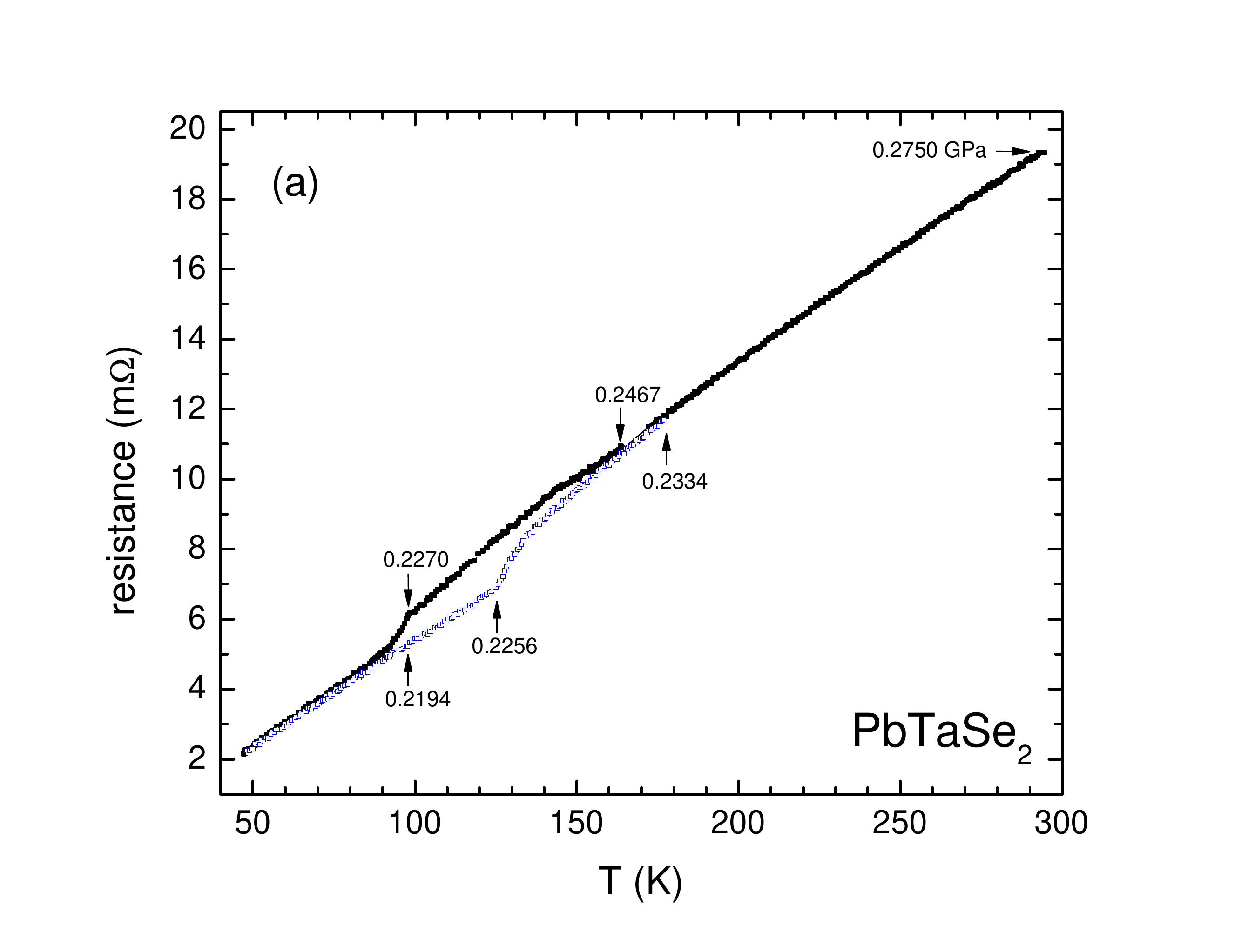}
\includegraphics[angle=0,width=100mm]{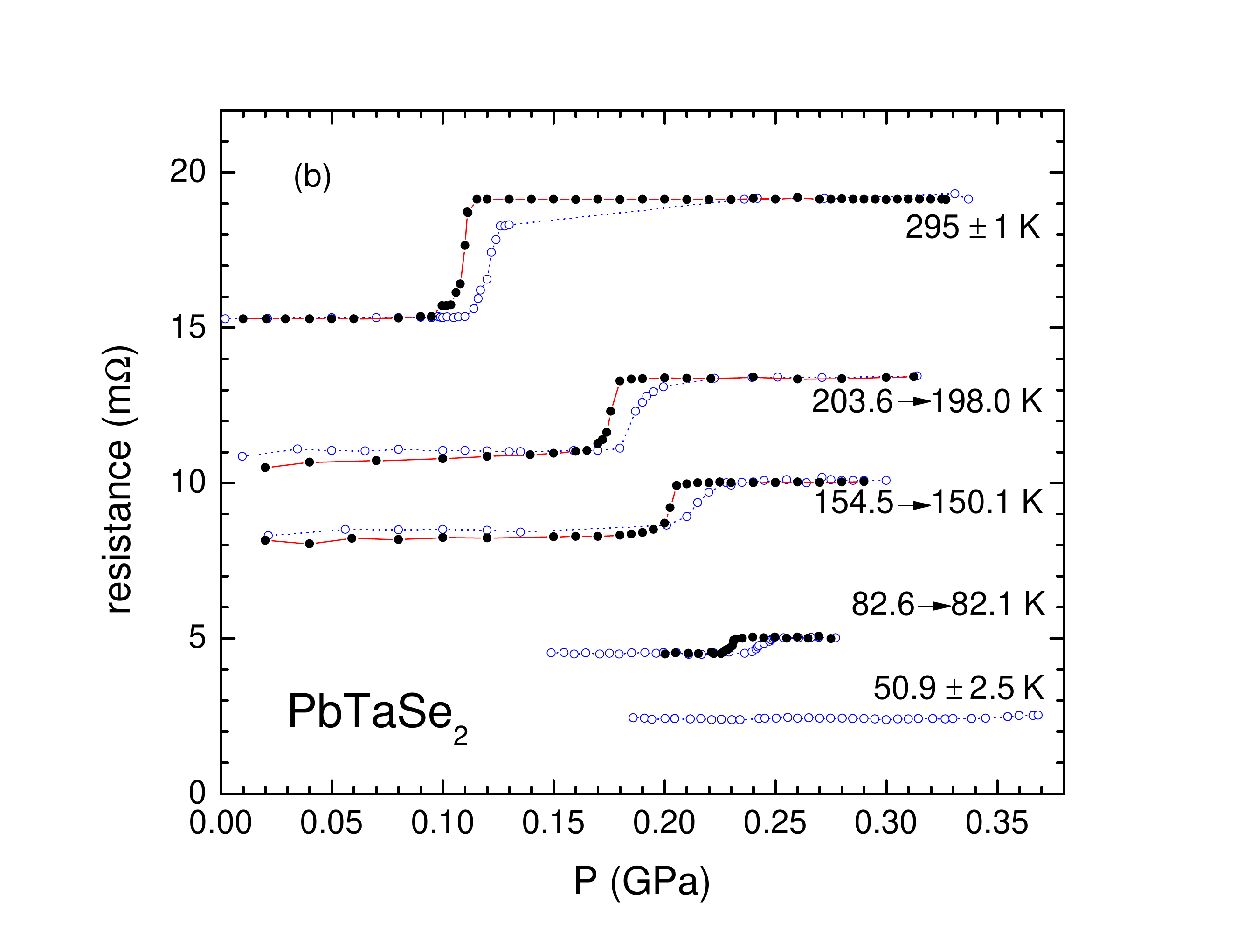}
\end{center}
\caption{(Color online) Examples (a) of  temperature sweeps at almost constant, continuously monitored, gas pressure, and (b) of pressure sweeps close to constant temperature. Numbers in panel (a) indicate measured pressure in GPa, in panel (b) change of temperature during the run. Arrows on the curves indicate the direction of the temperature/pressure changes: open symbols: increase, filled symbols - decrease. } \label{F10}
\end{figure}

\begin{figure}
\begin{center}
\includegraphics[angle=0,width=120mm]{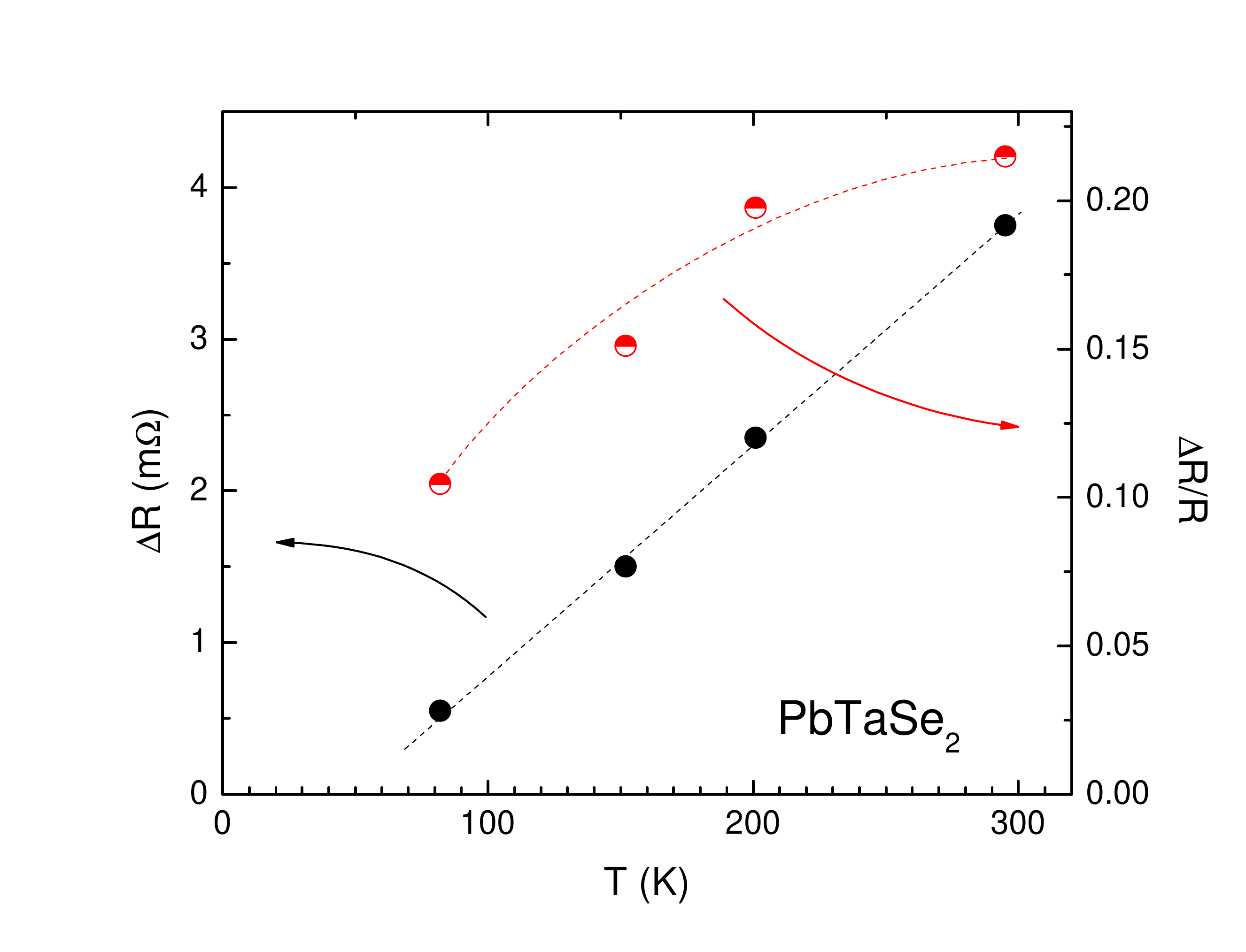}
\end{center}
\caption{(Color online) Temperature-dependent absolute and relative (to the average resistance below and above the transition) resistance jump at the apparent structural transition. Dashed lines are the guides for the eye.} \label{F11}
\end{figure}

To address the pressure dependence of the apparent structural transition in a quantitative manner, a set of measurements in a He gas system was performed (Fig. \ref{F10}). The hysteretic nature of the transition is seen both in temperature sweeps and pressure sweeps. These signatures are sharp and well defined. From the pressure sweeps (Fig. \ref{F10}(b))  it is seen that (a) at lower temperatures the transition shifts to higher pressures; (b) as has been seen in figure \ref{F9}, the size of the resistance jump becomes smaller at lower temperatures and it appears not to be detected anymore in the pressure sweep at $\sim 50$ K. Fig. \ref{F11} shows that both absolute and relative resistance jump decrease at lower temperatures, and it is not clear if at 50 K the jump disappears, or just becomes on the level of the noise in the data.

\begin{figure}
\begin{center}
\includegraphics[angle=0,width=120mm]{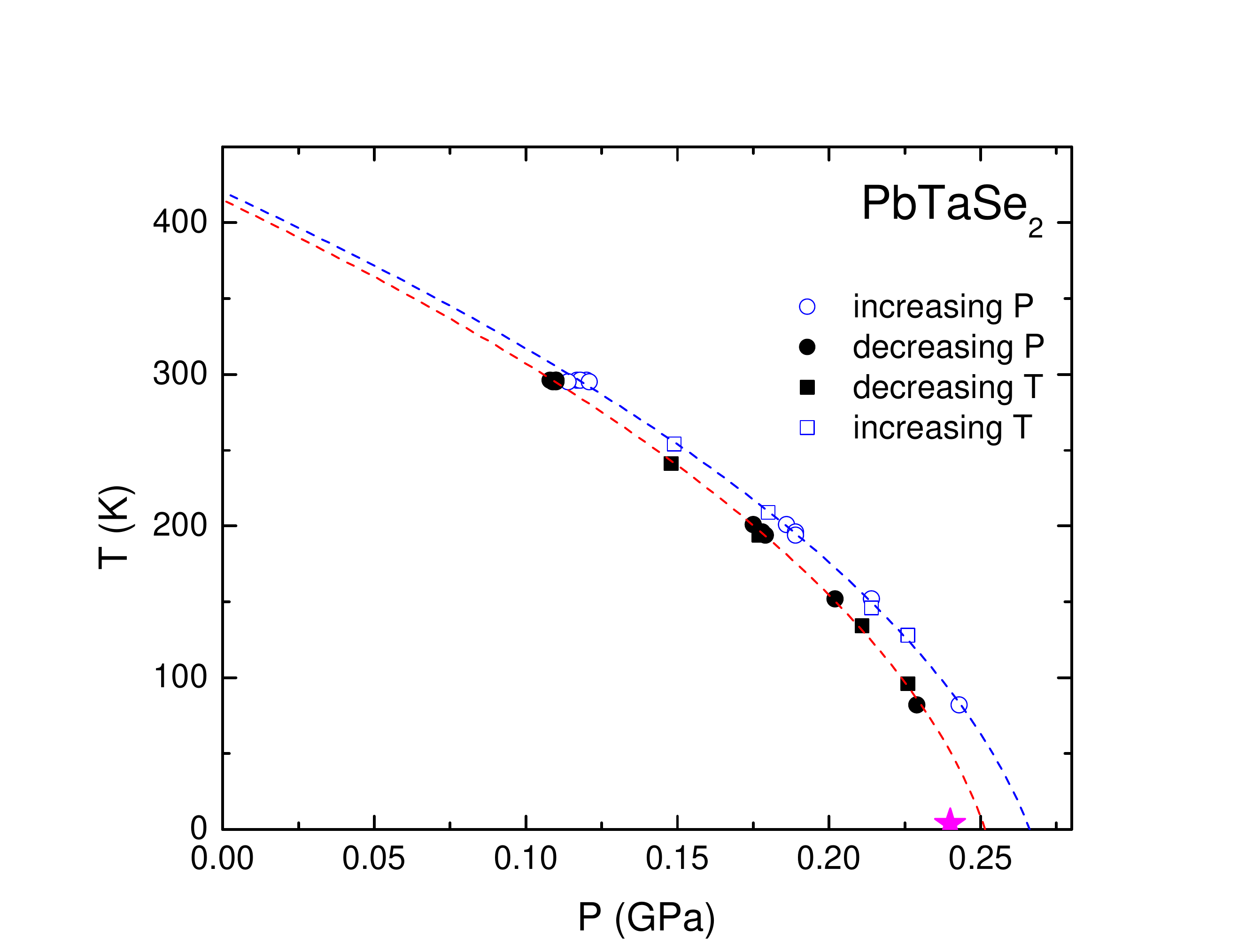}
\end{center}
\caption{(Color online) Pressure dependence of the apparent structural transition. Dashed lines are from the second order polynomial fits of the $P(T)$ data. Star - data point corresponding to two-step superconducting transition in Fig. \ref{F3}a.} \label{F12}
\end{figure}

The pressure dependence of the apparent structural transition measured in gas pressure system is presented in Fig. \ref{F12}. The second order polynomial fits to the $P(T)$ data extrapolate to 0.25 GPa for the decreasing temperature manifold to intercept the $T = 0$ K axis. In a similar manner, both manifolds extrapolate  to $\sim 425$ K at ambient pressure. This is a rather rough extrapolation, additional,  preferably structural, data at high temperatures are required to verify the nature of the transition.

\begin{figure}
\begin{center}
\includegraphics[angle=0,width=120mm]{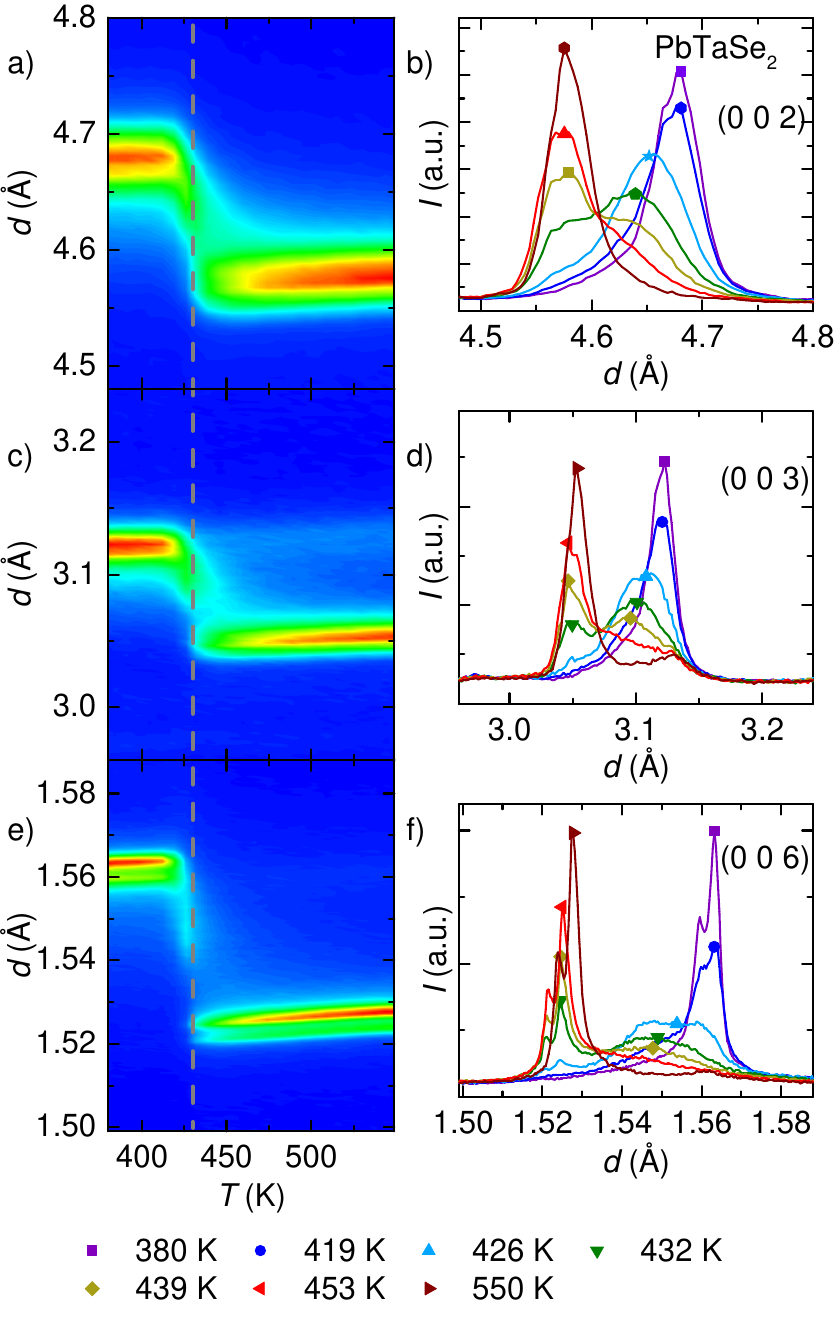}
\end{center}
\caption{(Color online) Evolution of the peaks corresponding to (0~0~2), (0~0~3), and (0~0~6) reflections at elevated temperatures (measured on heating) plotted as color maps of $d$-spacing vs. temperature [a), c), e)], and selected $2 \theta$ scans plotted as intensity vs. $d$-spacing  at different temperatures [b), d), f)].} \label{F15}
\end{figure}

In situ XRD was performed at different elevated temperatures. As mentioned above, the PbTaSe$_2$ powder retained the micaceous morphology, so only (00l) reflections could be detected and followed as a function of temperature. The experimental results are shown in Fig. \ref{F15}.  The (00l) reflections show a clear, step-like, contraction in the $c$-axis lattice on heating which is reversible upon cooling.  The temperature of this transition is $\sim 425-430$ K.

\begin{figure}
\begin{center}
\includegraphics[angle=0,width=120mm]{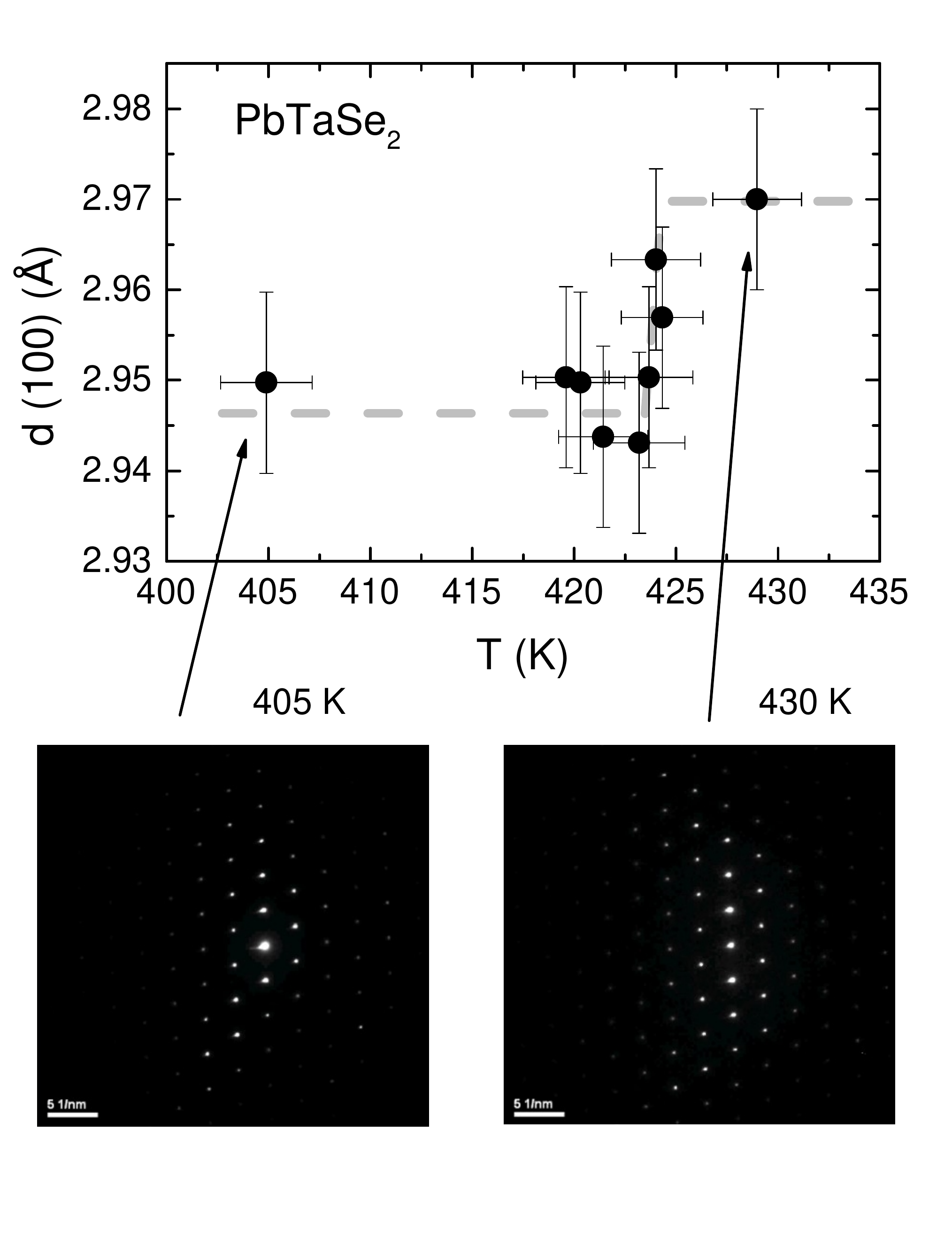}
\end{center}
\caption{In-plane TEM results at elevated temperatures. Upper panel: the basal plane $d$-spacing, $d (100)$ as a function of temperature. Lower panels: selected area diffraction patterns at 405 K and 430 K.  Gray dashed line is a guide to the eye.} \label{F16}
\end{figure}

To resolve changes of the in-plane lattice with heating, TEM measurements were performed (Fig. \ref{F16}).  The thin region studied with TEM was co-planar with the basal planes, providing an orthogonal view of the lattice expansion on heating compared to the XRD results, i.e., the (hk0) reflections.  Continuous recording of the selected area diffraction was obtained on cooling. The sample was slowly cooled over the temperature range of $\sim 430$ K to $\sim 400$ K at $\sim 13$ K/min. An abrupt contraction of the basal plane lattice of $\sim 0.03 \AA$ occurred $\sim 425$ K. No other changes in the diffraction pattern were observed, indicating that like in the XRD there is a discrete change in the cell parameters occurring at $\sim 425$ to 430 K but no obvious change in the space group (Fig. \ref{F16}, lower panels). TEM results suggest that the basal plane undergoes a normal expansion with heating and contraction on cooling in contrast to the XRD results which shows a large contraction in the $c$-axis at $\sim 425$ K.

\begin{figure}
\begin{center}
\includegraphics[angle=0,width=120mm]{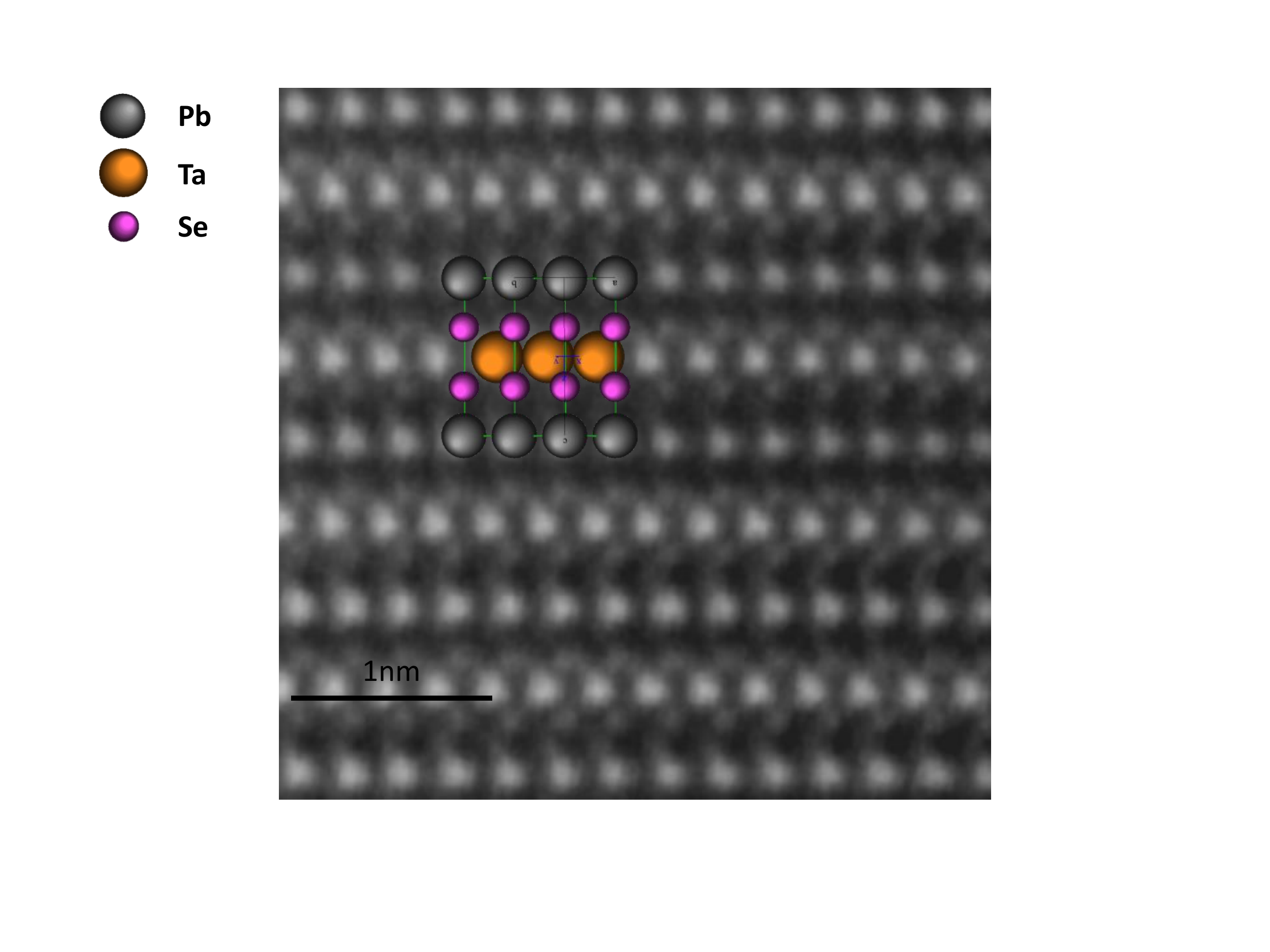}
\end{center}
\caption{(Color online) Atomic resolution TEM  imaging at room temperature. Note fairly large gap between the Ta/Se layer and Pb layer.} \label{F17}
\end{figure}

Aberration corrected scanning transmission electron microscopy using a FEI Titan Themis 300 Cubed 300 STEM/TEM shows that the room temperature atomic decoration is fully consistent with the space group \# 187, $P\bar{6}m2$, (Fig. \ref{F17}). The image suggest minimal chemical disorder. There is a rather large gap between the Pb layers relative to the Ta/Se inner layers which form an open network of edge sharing prisms.  This large gap between these layers maybe responsible for the lattice contraction in the $c$-axis with heating.

Altogether on heating through the structural transition at $\sim 425 - 430$ K  the $a$-axis increases by $\sim 0.8 \%$, whereas the $c$-axis decreases by $\sim 2 \%$, leading to a decrease of the unit cell volume by $\sim 0.4 \%$.

\section{Discussion and summary}

\begin{figure}
\begin{center}
\includegraphics[angle=0,width=120mm]{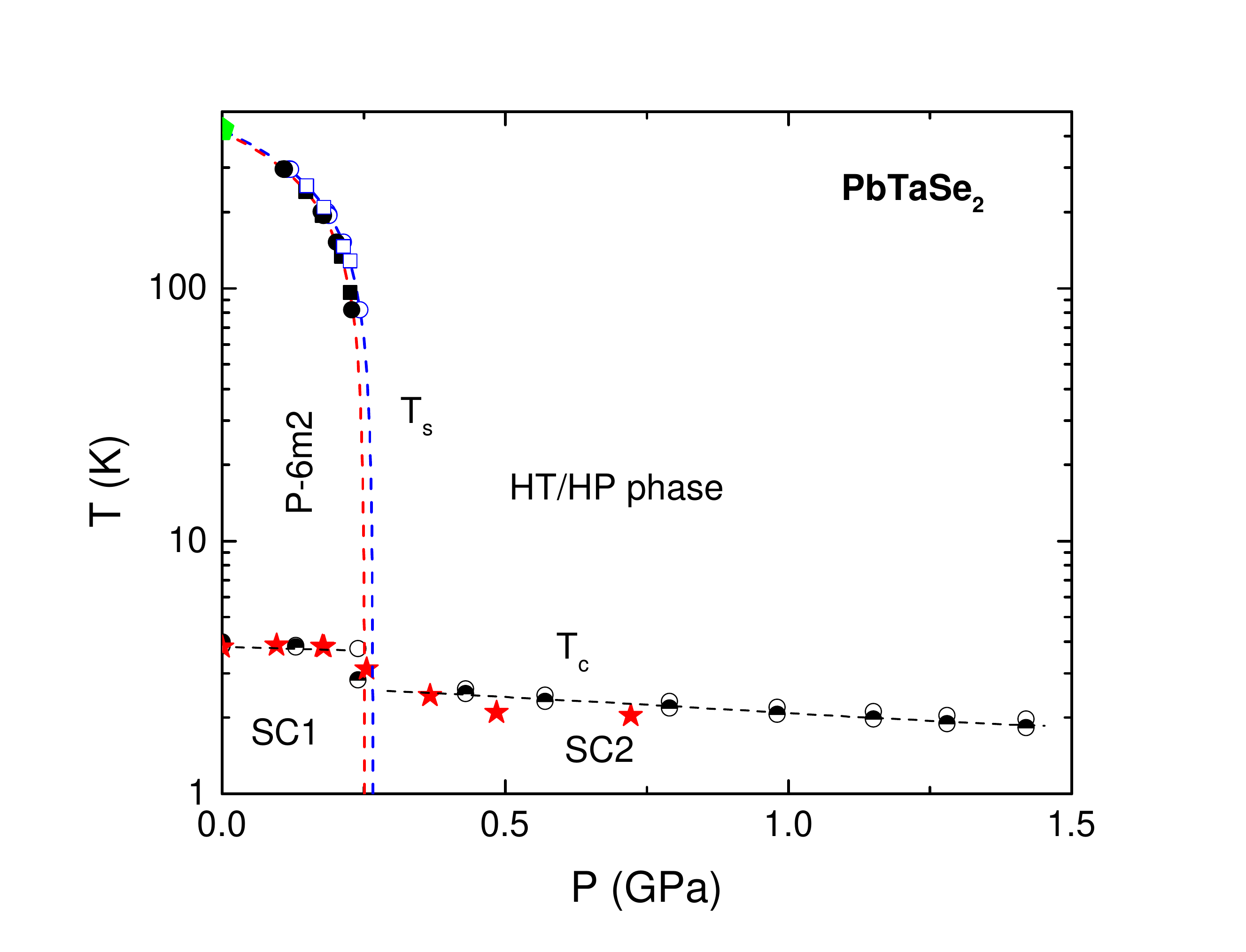}
\end{center}
\caption{(Color online) Pressure - temperature phase diagram on a semi-log scale.Two superconducting and two structural phases are labeled.  Lines are guides to the eye. HT / HP denotes high temperature / high pressure. Green pentagon - ambient pressure data point from high temperature TEM.} \label{F13}
\end{figure}

\begin{figure}
\begin{center}
\includegraphics[angle=0,width=120mm]{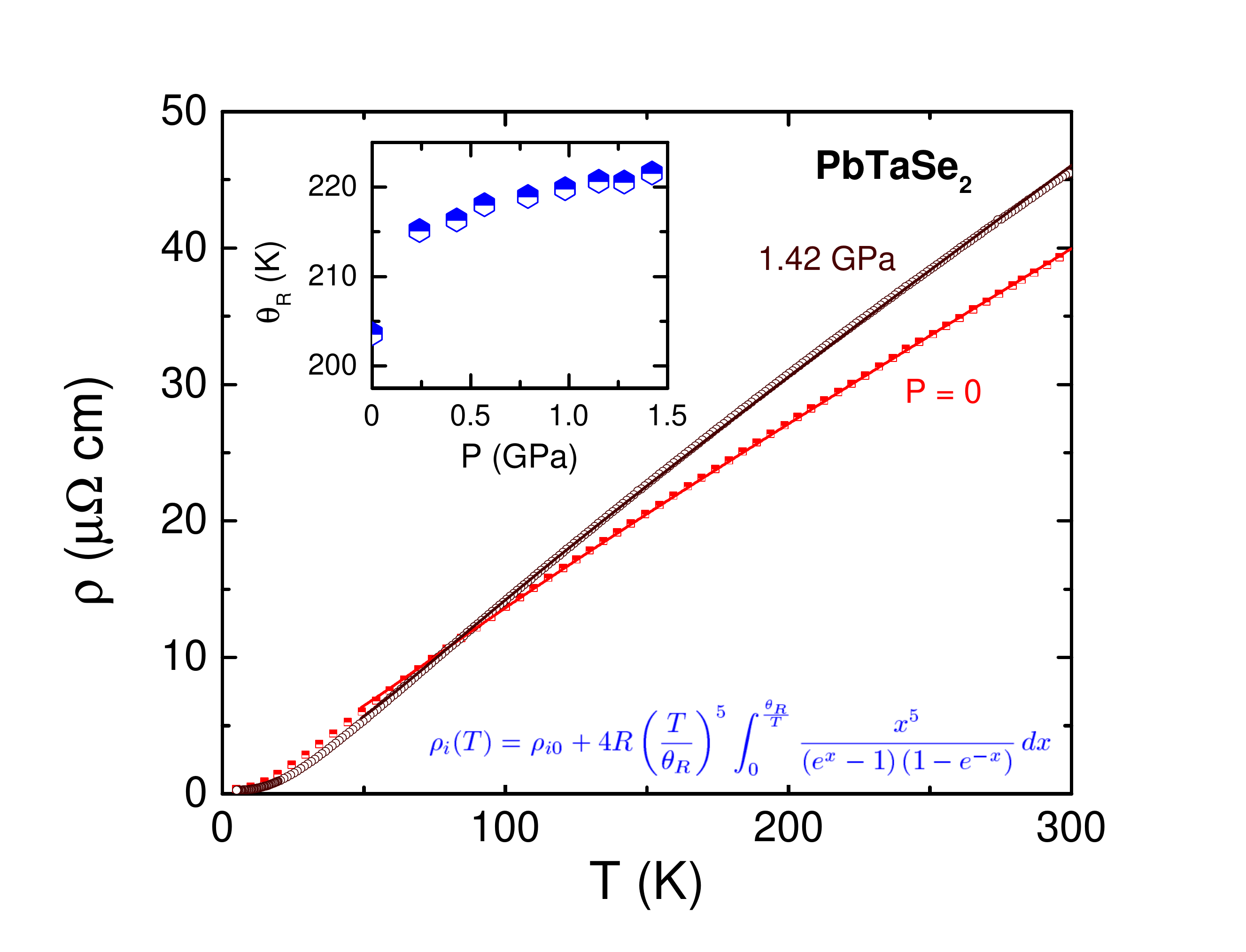}
\end{center}
\caption{(Color online) Examples of the Bloch - Gr\"uneisen fits of resistivity. Inset: Debye temperature as a function of the low temperature  pressure values (see text).} \label{F14}
\end{figure}

Based on the data discussed above, the pressure - temperature the phase diagram for PbTaSe$_2$ (Fig. \ref{F13})  appears to have two superconducting phases with the boundary between them defined by a structural phase transition that has extremely steep pressure dependence. Since the normal state resistivity, magnetoresistance and the initial slopes of $H_{c2}(P)$ are different in these two phases, clearly, the electronic structure is affected. A Bloch - Gr\"uneisen fit of resistivity (over the 50 K - 300 K temperature range) (fig. \ref{F14}). suggests that as a result of the structural phase transition the Debye temperature increases, so that the lattice becomes stiffer. This is largely consistent with an overall decrease of the unit cell volume at ambient pressure on heating  through the transition. Despite this, $T_c$ is lower in the new structural phase, suggesting that either the change in the electronic subsystem is the dominant contribution to the $T_c$ decrease, or that the  in-plane phonons are more important for superconductivity than the out-of-plane phonons. The latter hypothesis is in agreement with the observed large gaps between the layers (Fig. \ref{F17}).

Since available experimental techniques do not allow for an unambiguous identification of the high pressure (high temperature) phase, we have performed first-principles calculations that address relative stability of several related hexagonal phases under pressure. 

\begin{figure}
\begin{center}
\includegraphics[angle=0,width=120mm]{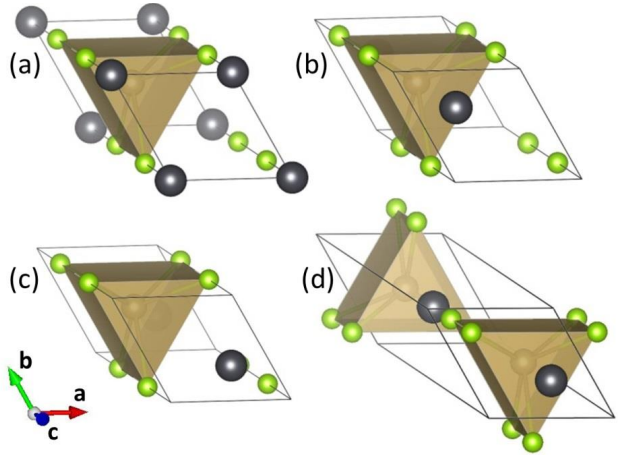}
\end{center}
\caption{(Color online) Crystal structures of (a) $P\bar{6}m2$, (b) $Pb{-}1c$, (c) $Pb{-}1e$ and (d) $hex2$  PbTaSe$_2$ structures. Dark grey and green balls are Pb and Se atoms. Shaded balls at centers of brown triangle prisms are Ta atoms.} \label{F18}
\end{figure}

In addition to the known $P\bar{6}m2$ structure of PbTaSe$_2$, we consider 3 other low-energy structures (within 50.0 meV/atom with respect to that of the $P\bar{6}m2$ structure) coming from our crystal structure optimization scheme with one and two formula units and hexagonal symmetry constraints (Fig. \ref{F18}). These structures can also be obtained through modification of the $P\bar{6}m2$ structure. The $Pb{-}1c$ structure can be obtained from the $P\bar{6}m2$ structure by moving the Pb atom from the $1a$-Wyckoff position to the $1c$-Wyckoff position. Similarly the $Pb{-}1e$ structure can be obtained by shifting the Pb from the $1a$-Wyckoff position  to the $1e$-Wyckoff position. By doubling unit cell of the $P\bar{6}m2$ structure along the $c$-axis lattice vector and then moving the upper half of unit cell by 1/3 along the long diagonal of the basal plane $(1/3(\bold{b} - \bold{a}))$, the $hex2$ structure can be obtained. The $Pb{-}1c$ and $Pb{-}1e$ structures are in $P\bar{6}m2$ space group symmetry whereas the $hex2$ structure is in $P6_3mc$ space group symmetry. The lattice parameters and Wyckoff positions of these structures are given in Table \ref{T1}. The lateral lattice constants of these 3 structures are similar to that of the $P\bar{6}m2$ structure, but their lattice constants along the $c$ direction ($c/2$ for the $hex2$ structure) are smaller than that of the $P\bar{6}m2$ structure by 0.534 \AA, 0.487 \AA, and 0.246 \AA~respectively, where the $Pb{-}1c$ structure has the shortest lattice parameter $c$. 

\begin{figure}
\begin{center}
\includegraphics[angle=0,width=120mm]{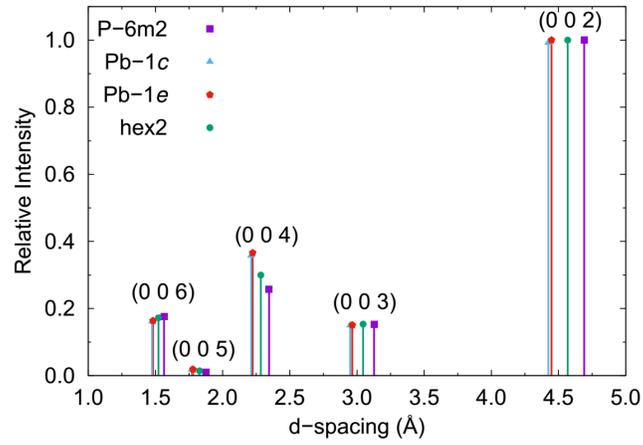}
\end{center}
\caption{(Color online) Simulated XRD patterns of different considered PbTaSe$_2$ structures (see text and Fig. \ref{F18} for details.)} \label{F19}
\end{figure}

In order to compare with experimental XRD  results, we simulated the XRD (0 0 l)-peaks of all four structures considered. It is interesting to note that although the $hex2$ structure has a doubled unit cell along the $c$-direction in comparison with other structures, it shows only (0 0 l)-peaks with even-l. As shown in Fig. \ref{F19}, the XRD (0 0 l)-peaks versus $d$-spacing patterns from all four structures are very similar. In figure \ref{F19} the peak indexes of the $hex2$ structure are (0 0 2l) of the labeled ones and the diffraction intensities of all structures are normalized by the intensity of their (0 0 2)-peak. 

\begin{figure}
\begin{center}
\includegraphics[angle=0,width=120mm]{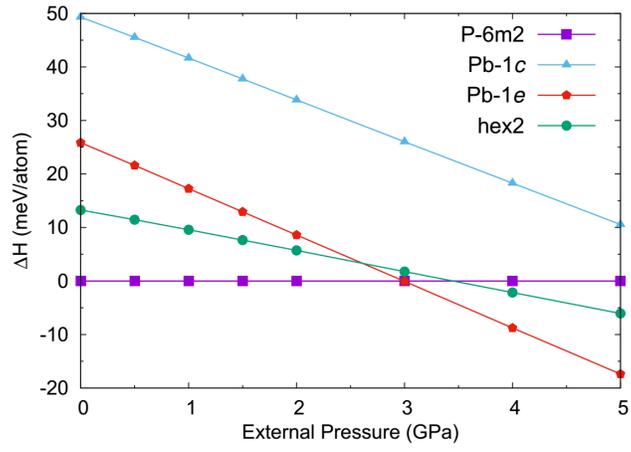}
\end{center}
\caption{(Color online) Formation enthalpy differences of the PbTaSe$_2$ structures under pressure with respect to the $P\bar{6}m2$ structure.} \label{F20}
\end{figure}

Figure \ref{F20} shows the relative formation enthalpies of different PbTaSe$_2$ structures as a function of pressure with respect to that of the $P\bar{6}m2$ structure. At zero and low pressure, the stable structure is the $P\bar{6}m2$ structure. As the pressure is increased, the $Pb{-}1e$ structure becomes more stable with a structural transition from the $P\bar{6}m2$ to the $Pb{-}1e$  at 3 GPa. At ambient conditions, the formation energy of the $Pb{-}1e$ structure is 25.9 meV/atom higher than that of the $P\bar{6}m2$ ground state structure. This energy difference is about the room temperature thermal energy. We note that the $hex2$ structure is lower in energy than the $Pb{-}1e$ structure at ambient conditions. However, the $Pb{-}1e$ structure becomes more stable than the $hex2$ structure when the external pressure is greater than 2.7 GPa, which is just below the transition pressure from the $P\bar{6}m2$ structure to the $Pb{-}1e$ structure. Thus these calculations suggest that the experimentally observed high pressure (high temperature) structure is the $Pb{-}1e$ structure. We would like to note that structural transition pressure from the DFT calculations is higher than that observed in experiment. This discrepancy is probably due to the systematic error in DFT calculation of pressure. For example, in the literature the predicted structural transition pressure of Si from cubic diamond (Si-I) to - Sn (Si-II) phase can be several GPa off the experimentally observed value, depending on exchange-correlation functional used.\cite{muj03a,hen10a}  But the sequence of phase transition, i.e. Si-I to Si-II to Si-V to Si-VI to Si-VII, from DFT calculation is consistent with experiment. \cite{muj03a}
\\

{\it To summarize},  PbTaSe$_2$ exhibits a structural, sharp, first order, phase transition at a very moderate pressure of $\sim 0.25$ GPa at low temperatures. PbTaSe$_2$ has a $\Delta T_s/\Delta P \approx - 1700$ K/GPa making it one of the more pressure sensitive transitions found  in inorganic compounds. The structural phase transition line extends to $\sim 425$ K at ambient pressure as evidenced by transmission electron microscopy and x-ray diffraction at elevated temperatures. Upon transition to the new phase (on increase of temperature at ambient pressure) the $c$-axis decreases and the $a$-axis increases, resulting in a slight, $\sim 0.4 \%$ decrease of the unit cell volume. The new, high temperature / high pressure phase has similar crystal structure.  As suggested by the first-principles calculations, in  this structure Pb shifts from $1a$  to $1e$ Wyckoff position with Ta and Se positions remaining the same.
The superconductivity appears to be robust, it persists through the structural phase transition into the other phase with slight, step-like decrease of $T_c$.

\appendix

\section{Ambient pressure upper critical field, $H_{c2}(T)$}

\begin{figure}
\begin{center}
\includegraphics[angle=0,width=80mm]{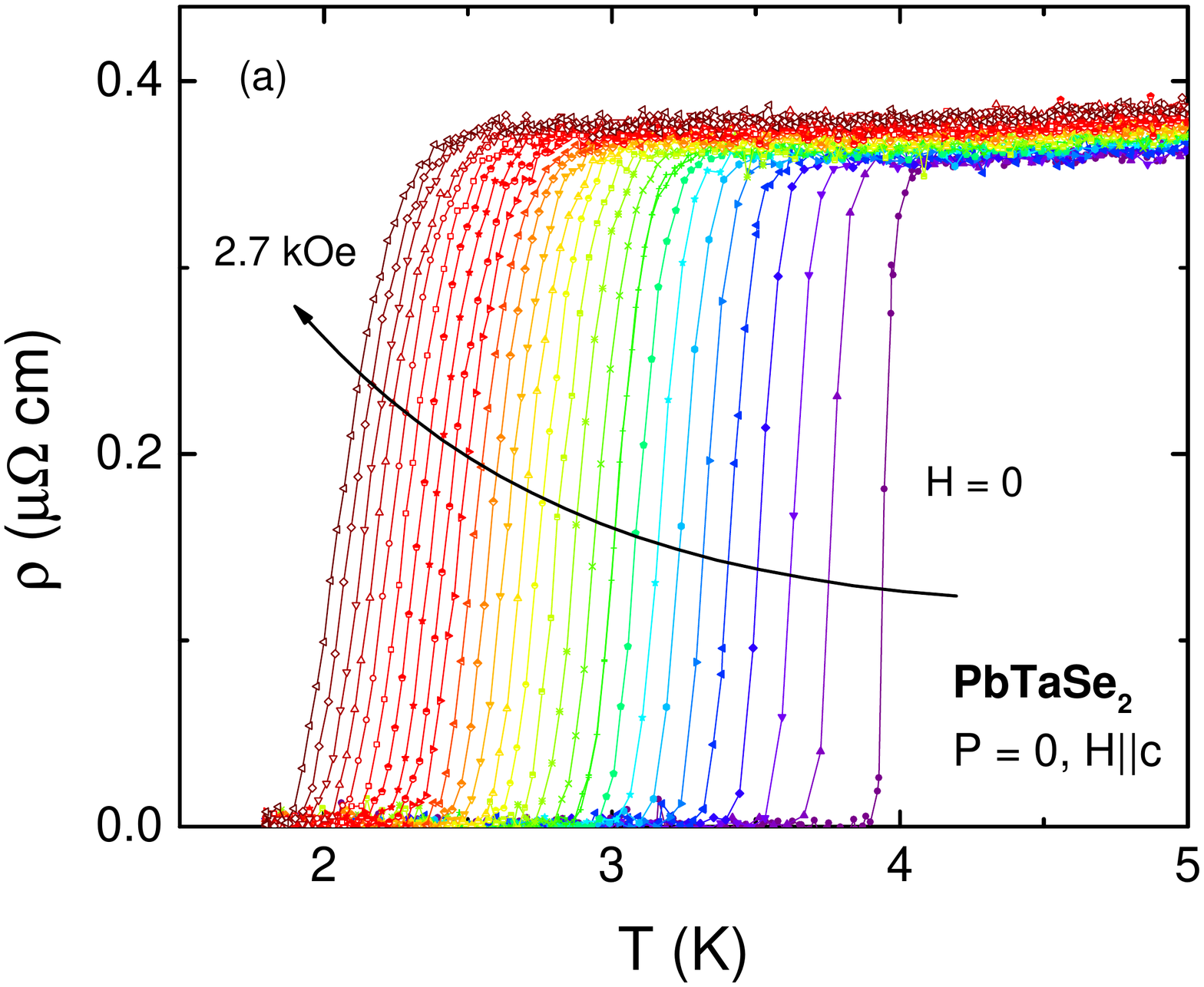}
\includegraphics[angle=0,width=80mm]{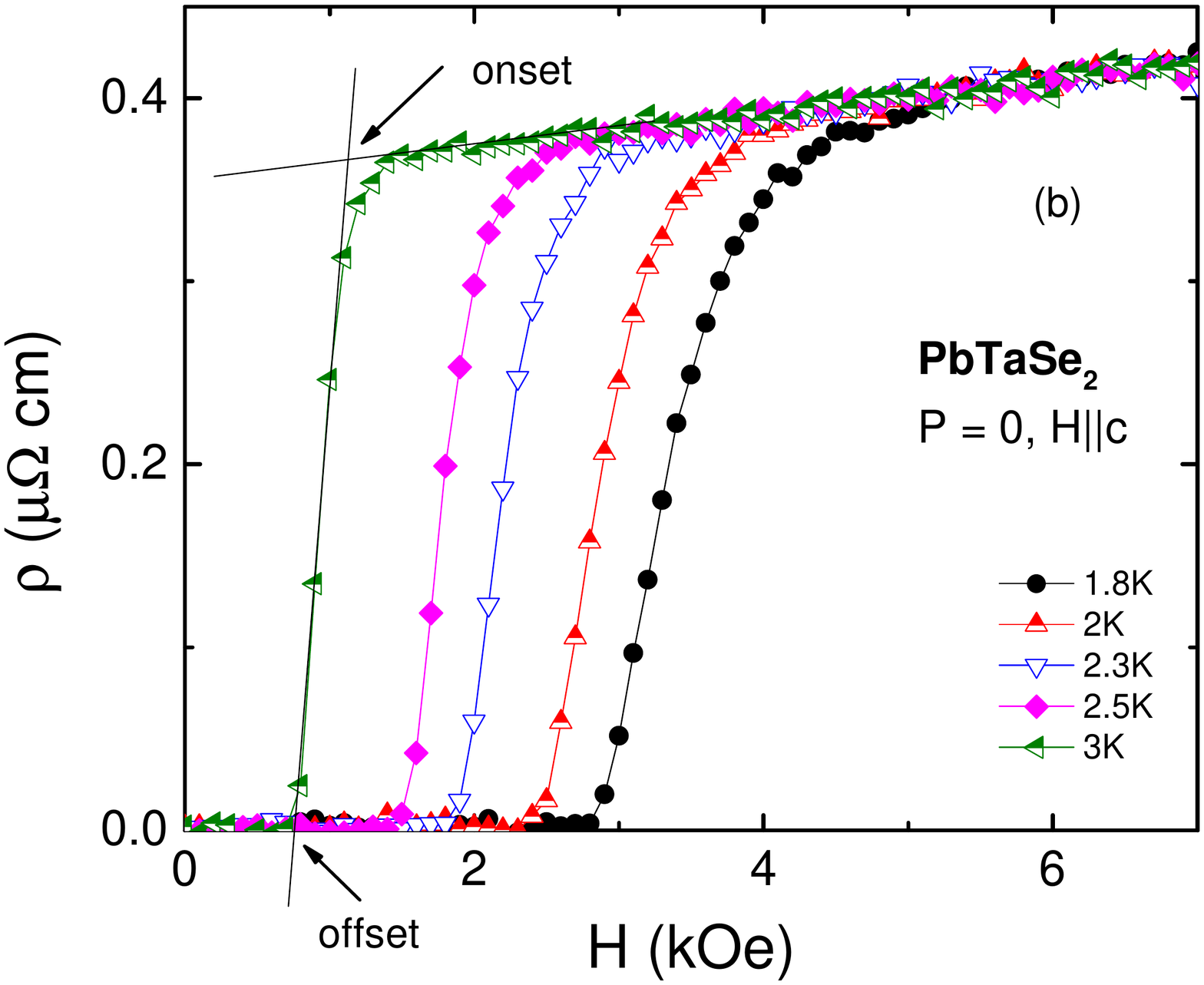}
\includegraphics[angle=0,width=80mm]{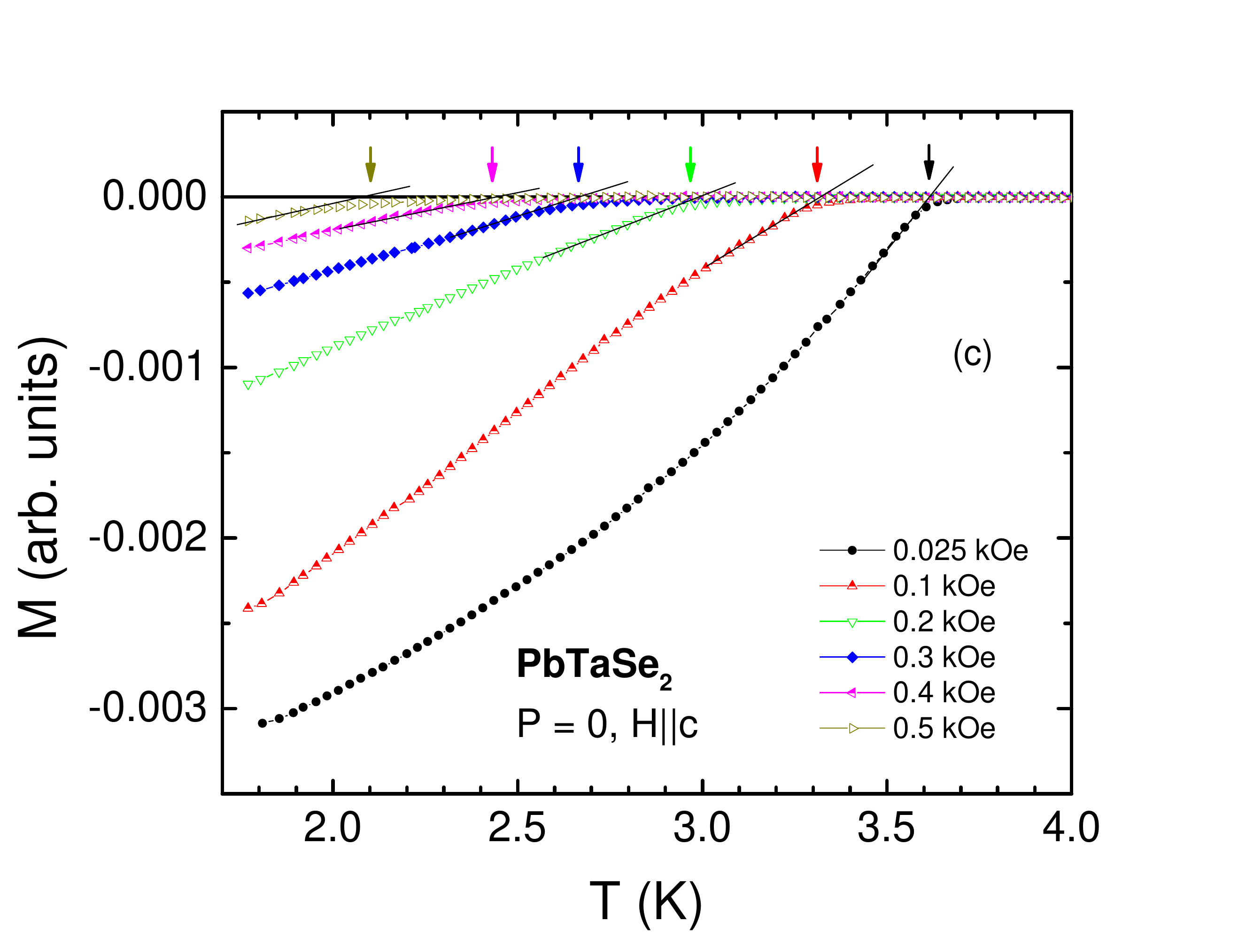}
\includegraphics[angle=0,width=80mm]{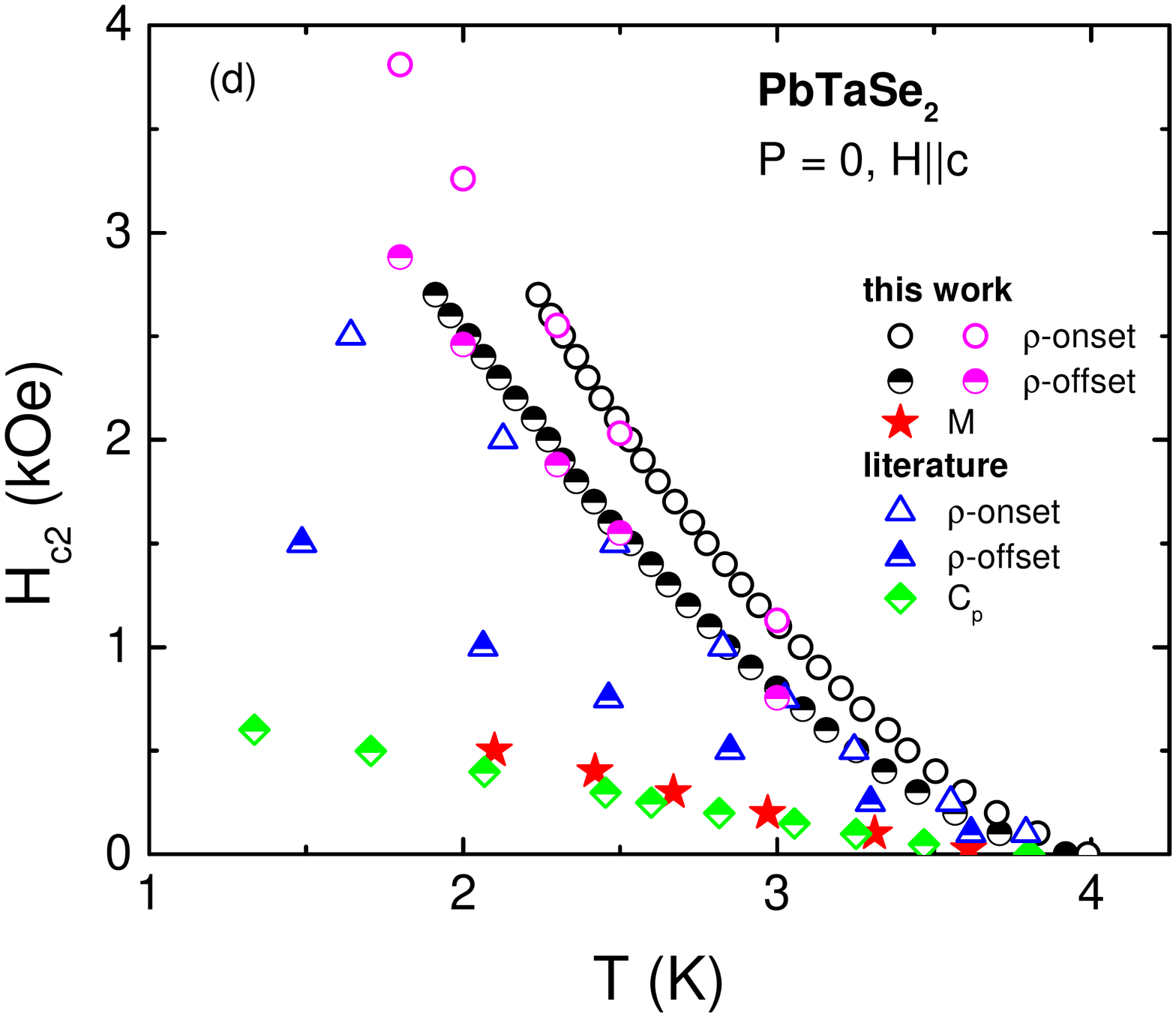}
\end{center}
\caption{(Color online) (a) Low temperature $\rho(T)$ data taken in applied magnetic fields from $H = 0$ to $H = 2.7$ kOe with 0.1 kOe steps. (b) Field dependent resistivity in fixed magnetic fields. Onset and offset criteria are shown for 3 K dataset. (c) Low temperature zero - field - cooled magnetization data taken in different magnetic fields. Arrows mark the superconducting transition temperatures. (d) Zero pressure $H_{c2}(T)$ data for $H \| c$ obtained from resistivity (onset and offset criteria) - red circles $\rho(H)$, black circles $\rho(T)$ , and magnetic susceptibility data - stars, plotted together with the literature data \cite{zha16a} from resistivity (triangles) and  heat capacity (rhombi) measurements in magnetic field.} \label{F2}
\end{figure}

Ambient pressure, low temperature resistivity, $\rho(T)$, measured  in  constant fields and $\rho(H)$, measured at  constant temperatures, are shown in Figs. \ref{F2}a,b. Two criteria, onset and offset (see Fig. \ref{F2}b) were used to evaluate the $H_{c2}$. Additionally, zero-field-cooled magnetization was measured in different applied fields (Fig. \ref{F2}c) and $H_{c2}(T)$ was determined from these measurements as well. The summary of these results, the ambient pressure $H_{c2}(T)$ for $H \| c$ inferred from electrical transport and magnetization measurements are shown in Fig. \ref{F2}d and are  compared to the literature data \cite{zha16a}. The resistively determined  $H_{c2}(T)$ is comparable with the literature data (with possible contributions to the difference from the sample shape and slight mis-orientation). Similarly, results from magnetization are very close to the published upper critical field determined from specific heat. However, both in the literature \cite{zha16a} and in this work, even though the low field / zero field $T_c$ values are quite similar, $H_{c2}(T)$ from electrical transport [$\rho(T,H)$] and thermodynamic [$M(T,H), C_p(T,H)$] measurements are noticeably different. The origin of this difference is not clear at this point.

\section{Resistive upper critical field under pressure}

\begin{figure}
\begin{center}
\includegraphics[angle=0,width=120mm]{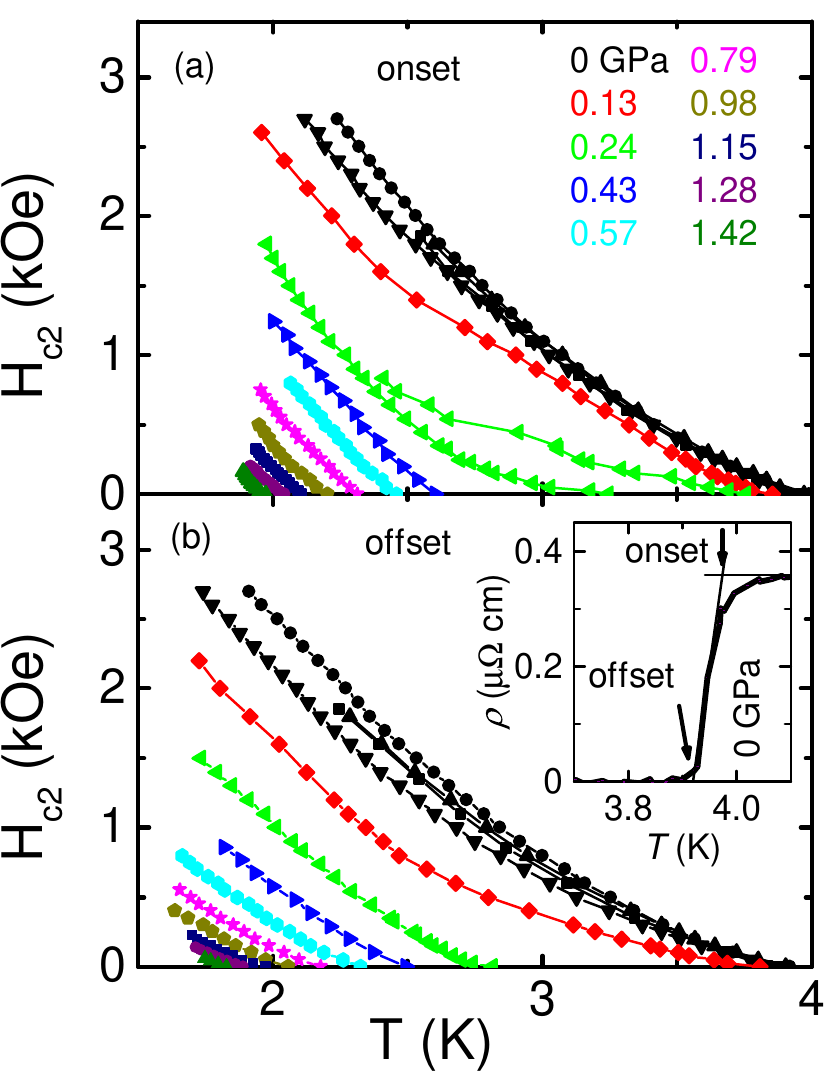}
\end{center}
\caption{(Color online) $H_{c2}(T)$ data for $H \| c$ obtained from resistivity (onset and offset criteria) at different pressures. Inset shows the definition of the criteria used.} \label{F5}
\end{figure}

\begin{figure}
\begin{center}
\includegraphics[angle=0,width=120mm]{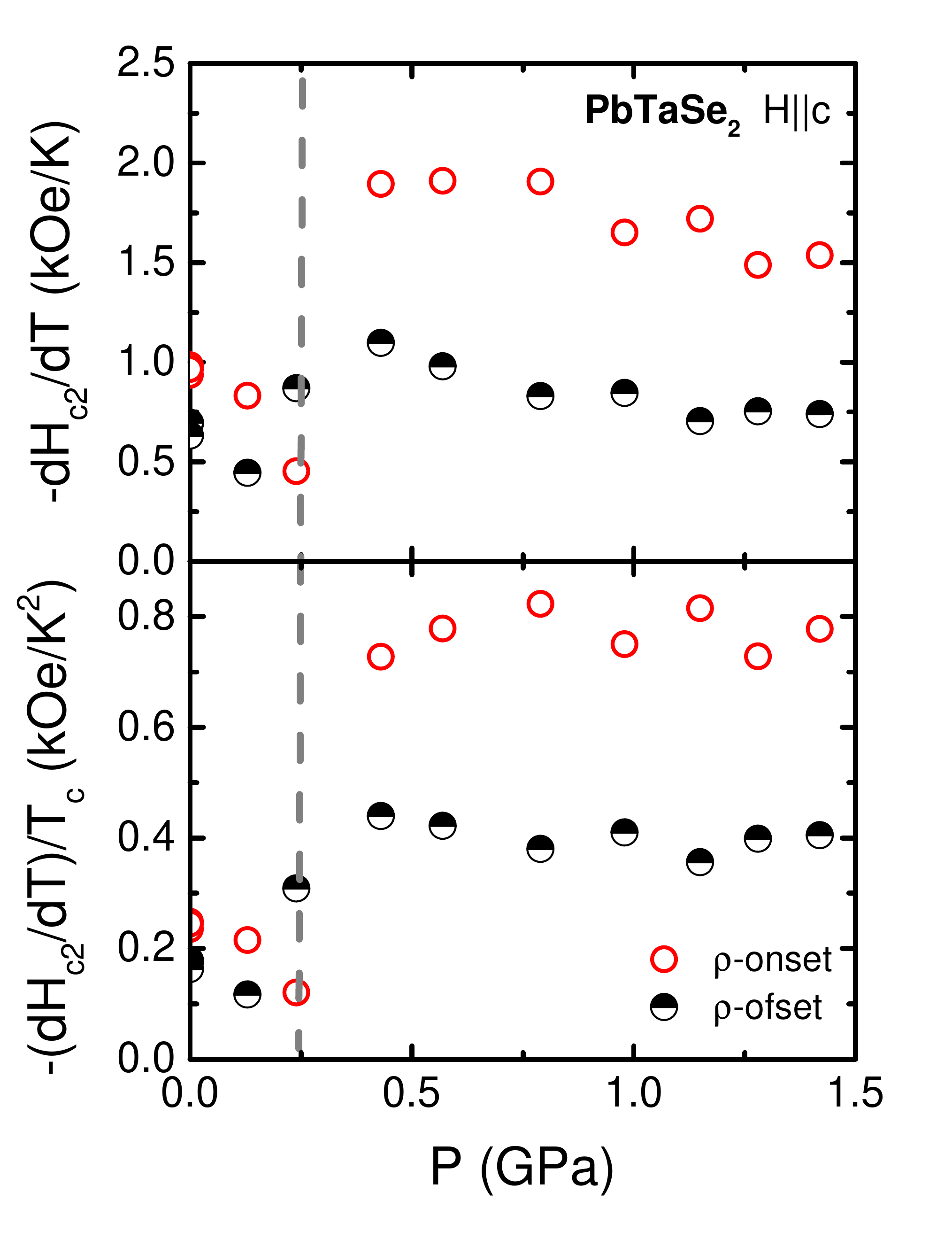}
\end{center}
\caption{(Color online) Pressure dependence of the initial, close to $T_c(H=0)$,  slope of the upper critical field for $H \| c$, plotted as $-dH{c2}/dT(P)$ and  $-(dH{c2}/dT)/T_c(P)$} \label{F6}
\end{figure}

The temperature-dependent upper critical field for $H \| c$ was measured resistively as a function of pressure (Fig. \ref{F5}). Whereas there is an apparent difference between thermodynamic and transport $H_{c2}$ values (see above), as well as an  upward curvature of $H_{c2}(T)$ and the limited range of the data for the higher pressures, we can still compare the change of the initial, close to $T_c(H=0)$, slope of transport $H_{c2}$ as a function of pressure (Fig. \ref{F6}). In agreement with the $T_c(P)$ behavior, both bare, $dH_{c2}/dT$, and normalized, $(dH_{c2}/ dT)T_c$,  initial slopes of the upper critical field have step-like change at $P \approx 0.25$ GPa. 

\begin{acknowledgments}

We thank the  Ms. Junying Shen for assistance in the He-gas measurements. Work at the Ames Laboratory was supported by the U.S. Department of Energy, Office of Science, Basic Energy Sciences, Materials Sciences and Engineering Division. The Ames Laboratory is operated for the U.S. Department of Energy by Iowa State University under contract No. DE-AC02-07CH11358, including a grant of computer time at the National Energy Research Scientific Computing Centre (NERSC) in Berkeley, CA. The work at Washington University was supported by the National Science Foundation (NSF) through Grants No. DMR-1104742 and 1505345.

\end{acknowledgments}

\clearpage

\begin{table}

\caption{Crystallographic data of PbTeSe$_2$ structures} \label{T1}

\begin{tabular}{cccccccc}

\hline\hline
{Structure}&{a(\AA)}&{c(\AA)}&\multicolumn{5}{c}{Wyckoff positions} \\
\hline\hline

& & &~Pb~&~$1a$~&~0.0~&~0.0~&~0.0 \\
$P\bar{6}m2$~&~3.415~&~9.382~&~Ta~&~$1d$~&~1/3~& ~2/3~&~1/2 \\
& & &~Se~&~$2g$~&~0.0~&~0.0~&~0.32283 \\
\hline

& & &~Pb~&~$2b$~&~1/3~&~2/3~&~0.49038 \\
$hex2$~&~3.419~&~18.237~&~Ta~&~$2b$~&~1/3~& ~2/3~&~0.24142 \\
& & &~Se~&~$2b$~&~1/3~&~2/3~&~0.83300 \\
& & &~Se~&~$2b$~&~1/3~&~2/3~&~0.65080 \\
\hline

& & &~Pb~&~$1e$~&~2/3~&~1/3~&~0.0 \\
$Pb{-}1e$~&~3.404~&~8.895~&~Ta~&~$1d$~&~1/3~& ~2/3~&~1/2 \\
& & &~Se~&~$2g$~&~0.0~&~0.0~&~0.31109 \\
\hline

& & &~Pb~&~$1c$~&~1/3~&~2/3~&~0.0 \\
$Pb{-}1c$~&~3.426~&~8.848~&~Ta~&~$1d$~&~1/3~& ~2/3~&~1/2 \\
& & &~Se~&~$2g$~&~0.0~&~0.0~&~0.31147 \\
\hline

\end{tabular}

\end{table} 

\clearpage

\clearpage

\clearpage

\clearpage

\clearpage

\clearpage

\clearpage

\clearpage

\clearpage

\clearpage

\clearpage

\clearpage

\clearpage

\clearpage

\clearpage

\clearpage

\clearpage

\clearpage

\clearpage

\clearpage

\clearpage

\end{document}